\documentclass[aps,prd,superscriptaddress,nofootinbib,eqsecnum,twocolumn]{revtex4}

\pdfoutput=1

\usepackage{graphicx}   
\usepackage{subfigure}
\usepackage{hyperref}
\usepackage{color}
\usepackage[dvipsnames]{xcolor}
\usepackage{amsmath,amsfonts,amsthm} 
\usepackage{bbold}
\usepackage{slashed}
\usepackage{soul}
\allowdisplaybreaks

\begin{document}

\title{Formulating the Kramers problem in field theory}

\author{Arjun Berera} 
\affiliation{School of Physics and Astronomy, 
University of Edinburgh, Edinburgh, EH9 3FD, United Kingdom}

\author{Jo\"el Mabillard} 
\affiliation{School of Physics and Astronomy, 
University of Edinburgh, Edinburgh, EH9 3FD, United Kingdom}

\author{Bruno W. Mintz} 
\affiliation{Departamento de F\'{\i}sica Te\'orica, Universidade do
Estado do Rio de Janeiro, 20550-013 Rio de Janeiro, RJ, Brazil}

\author{Rudnei O. Ramos}  \affiliation{Departamento de F\'{\i}sica
  Te\'orica, Universidade do Estado do Rio de Janeiro, 20550-013 Rio
  de Janeiro, RJ, Brazil}


\begin{abstract} 

The escape problem is defined in the context of quantum field
theory. The escape rate is explicitly
derived for a scalar field governed by fluctuation-dissipation dynamics, 
through generalizing the  standard Kramers problem.
In the presence of thermal fluctuations, there is a  nonvanishing
probability for a classical background field to  escape
from the well. Different from nucleation or quantum tunneling processes, the escape problem does not require the minimum of the potential, where the field is initially located in a homogeneous configuration, to be a false vacuum. The simple and well-known related
problem of the escape of a classical  point particle due to random
forces is first reviewed. We then discuss the difficulties associated with a
well-defined formulation  of an escape rate for a scalar field and how
these can be overcome. A definition of the  Kramers
problem for a scalar field  and  a method to obtain the rate are provided.
Finally, we discuss some of the potential  applications of our results,
which can range from condensed matter systems, i.e.,
nonrelativistic fields, to applications in high-energy physics, like
for cosmological phase transitions.

\end{abstract}


\maketitle

\section{Introduction}
 
The problem of escaping a potential well has been an active field of
research over the last century and has  applications in several
scientific disciplines, such as in physics and chemistry. Classically,
a particle put at rest at the bottom of a potential well stays there
if left undisturbed. However, in any realistic physical system, we
expect the presence of fluctuation and dissipation dynamics,  which,
for example, naturally emerge through the interactions  of the system
with a thermal bath. Under these conditions, an escape from the
potential well might be allowed.  The derivation of the escape rate is
called the Kramers problem~\cite{KRAMERS1940284} and is, to a large extent, well understood
for the  simplest systems, such as a classical point particle. One should note that, in its original zero-dimensional (usual field theory nomenclature for a point particle as a zero space and one time 
dimensional field) formulation, the escape problem is defined regardless of what is beyond the top of the energy barrier. That is, one is interested in the probability per unit time for the particle to escape the potential well, independently of what happens after this escape. To
our knowledge, however,  no explicit extension of this problem to a
relativistic field has been done so far.  Such an extension would be
very welcome for several possible applications in high-energy physics
and cosmology. {}For example, in the physics of the early
Universe, one is interested in describing cosmological fields immersed
in a hot medium. Fluctuation-dissipation dynamics has been shown to have interesting applications in the early Universe such as the
warm inflation paradigm \cite{Berera:1995ie,Berera:1999ws} and during
a cosmological phase transition \cite{Bartrum:2014fla}. 
It, therefore, follows to try to define and understand precisely 
the rate of escape of such fields due to thermal
fluctuations. 
 
Computing the probability for a classical particle to diffuse has
always been of great interest,  in particular in the context of
stochastic dynamics. Several methods have been proposed over the
years.  Kramers, a pioneer in the field, derived the so-called
Kramers rate~\cite{KRAMERS1940284} using the flux-over-population
method based on ideas originally developed by Farkas in
Ref.~\cite{L.:1927aa}. Another way to obtain the escape rate is
achieved with the mean-first-passage-time (MFPT) formalism using the
adjoint Fokker-Planck (FP)
operator~\cite{doi:10.1137/0133024,Talkner1987}. However, this
approach is more delicate to handle due to complex boundary
conditions.  A third method consists of finding the smallest positive,
nonvanishing, eigenvalue of the FP operator.  It has been shown that
this eigenvalue is directly related to the escape
rate~\cite{risken1996fokker}.   A comprehensive review of these
methods can be found in ~\cite{RevModPhys.62.251}. More recently,
in ~\cite{PhysRevE.60.R1}, it was shown that there is a universal equivalence
between these different approaches.

When regarding a field instead of a particle, the situation changes
significantly. A somewhat related problem in quantum physics is that
of the study of quantum tunneling.  The decay rate of a field has been
derived by Callan and Coleman at zero
temperature~\cite{Coleman:1977py,Callan:1977pt}  and extended to
finite temperature by Linde~\cite{LINDE1983421} (also known as
the nucleation problem in finite temperature quantum field
theory~\cite{PhysRevD.46.1379}). The inclusion of gravitational effects has been  studied by
Coleman and de Luccia in Ref.~\cite{Coleman:1980aw}.  The
formalism describing a field subject to random forces is
a developed topic
called stochastic
field theory ~\cite{Parisi:1988nd,zinn2002quantum}, 
but in spite of this there has never been a precise
and complete discussion of the escape problem.
One of the main difficulties is the identification of the
most suitable approach to be generalized to  a scalar
field. Zinn-Justin in Ref.~\cite{zinn2002quantum}  briefly states
the problem and suggests deriving the smallest eigenvalue  and the use
of instantons. This is indeed a possibility but, unfortunately, it
faces some analytical limitations when deriving the  rate. 
We find that the work of
Langer~\cite{LANGER1967108,LANGER1969258} in extending the
flux-over-population method to a $2N$-dimensional  system appears as
the most promising approach to be used with a field.

The field theory aspect of the problem renders the definition of an
escape more difficult and less intuitive than for a single point
particle.  In particular, the actual shape of the potential beyond the potential well plays a role in the computation of the rate for the field. However, as in the zero-dimensional case, the Kramers problem can be defined for both an initial true or false vacuum. Using the ideas and the formalism of the flux-over-population method extended to a field, we will propose in
this work a definition of the  Kramers problem and explicitly evaluate
the rate. Along the way in this derivation, we will encounter some familiar
situations, such as the Hawking  and Moss
solution~\cite{HAWKING198235}. We will also compare our final result
for the escape rate with the known result of nucleation rate due to thermal fluctuations~\cite{LINDE1983421,PhysRevD.46.1379}. In
particular,  considering the well-known result of Linde for the quantum tunneling rate at finite temperature~\cite{LINDE1983421}, we will show that, in the limit where the temperature is sufficiently high for the thermal fluctuations to dominate over the quantum fluctuations, the nucleation rate is proportional to the escape rate. This is remarkable since the two results are based on completely different approaches. This result shows that, when the system is initially in a false vacuum, the nucleation rate is indeed a special case of the escape rate.

Apart from the formal interest in the computation of an escape rate
for a scalar field, the result has potentially many applications.
The process helps to give a thorough understanding of
out-of-equilibrium situations, for example during phase
transitions. In particular, this process can influence the formation of
topological defects and potentially alter the stability of the
embedded configurations. In addition, the escape rate is a well-suited
mechanism for situations where the field needs to probe several local
minima. Such a situation appears in string theory, with the string
landscape, and, also, in condensed matter physics, in the context of
the glass transition, just to cite some of the potential
applications.  Another interesting application can be to stochastic inflation. Moreover, our derivation is formally identical to the
stochastic  quantization,  used in particular in lattice field theory,
where the origin of the stochastic forces is quantum instead of
thermal. A  precise knowledge of a transition rate is therefore of
great interest in this context.

The aim of this paper is, therefore, to formulate a
known problem, the definition and the derivation of the
Kramers escape rate, to a physical situation where it has
not been applied yet, a scalar field theory.
For this, we first need to define the escape problem consistently in
the context of field theory.  We then will obtain
an explicit expression for the escape rate in field
theory.  For our derivation we have identified
an already known method, the flux-over-population method~\cite{KRAMERS1940284,LANGER1967108,LANGER1969258},
which we show can be generalized to attain the result
we seek.  The main result for the escape rate in
field theory is Eq.~\eqref{eq:finalresultescaperate}.  Unsurprisingly, it has many
similarities to the expression for the particle case,
but there are also distinct differences, as will be evident from
our derivation and summarized below this main result.
Once a general expression
for the rate is obtained, we also discuss some techniques,
such as the thin-wall approximation, in order to obtain
an analytical expression for the rate, when a potential
is specified. Beyond the actual result, the method itself
suggests an approach for several applications which is,
after all, as essential as the actual result for the rate.

The present paper is organized as follows.
Section~\ref{sec:classicalpointparticlereview} gives a brief review of the
Kramers problem and  the methods necessary to compute the escape rate
in the simplest case of a point particle. We will focus on one of the main approaches for the computation of
the Kramers rate, the flux-over-population method, which is the best candidate to be generalized to a scalar
field. Another approach, the MFPT, not used here, is however discussed in the  Appendix. The MFPT  provides a simple interpretation of the escape  rate as we will
see. We also present the proof of the equivalence of the two
methods.  Section~\ref{sec:escaperateforscalarfieldderivation}
first states the difficulties in the formulation of an escape problem
for a scalar field. We then review some basics of stochastic field
theory with the  Langevin and the associated FP equation. We then
define and derive the escape rate for a scalar field using the
flux-over-population  method. This is the main result of this work. In
Sec.~\ref{discussion} we compare our result for the escape rate with a
closely related problem, that of quantum tunneling dominated by
thermal fluctuations and outline some of the similarities and
fundamental differences between the two cases.
Section~\ref{sec:applicationsforcosmologyandbeyond} discusses some
potential applications for  our results. {}Finally,
Sec.~\ref{conclusions} has some concluding remarks.

 \section{Rate of Escape of a Classical Point Particle}
 \label{sec:classicalpointparticlereview}
 
 To introduce the escape problem and the associated computations, we
 consider the simplest and well-known example of a classical point
 particle  in a metastable potential, whose dynamics is subject to
 both a dissipative and a stochastic force.  The two
 equivalent formalisms, based on the Langevin and the FP equations,
are reviewed. We
 then also review the flux-over-population and the
 mean-first-passage-time methods, which are used to obtain the escape
 rate, that is, the probability per unit time that the particle
 crosses the top of the potential energy barrier (and, therefore,
 escapes  the metastable minimum). 

 \subsection{Point particle in a metastable potential}

\begin{figure}[htb!]\centering
 {\includegraphics[width=0.75\columnwidth]{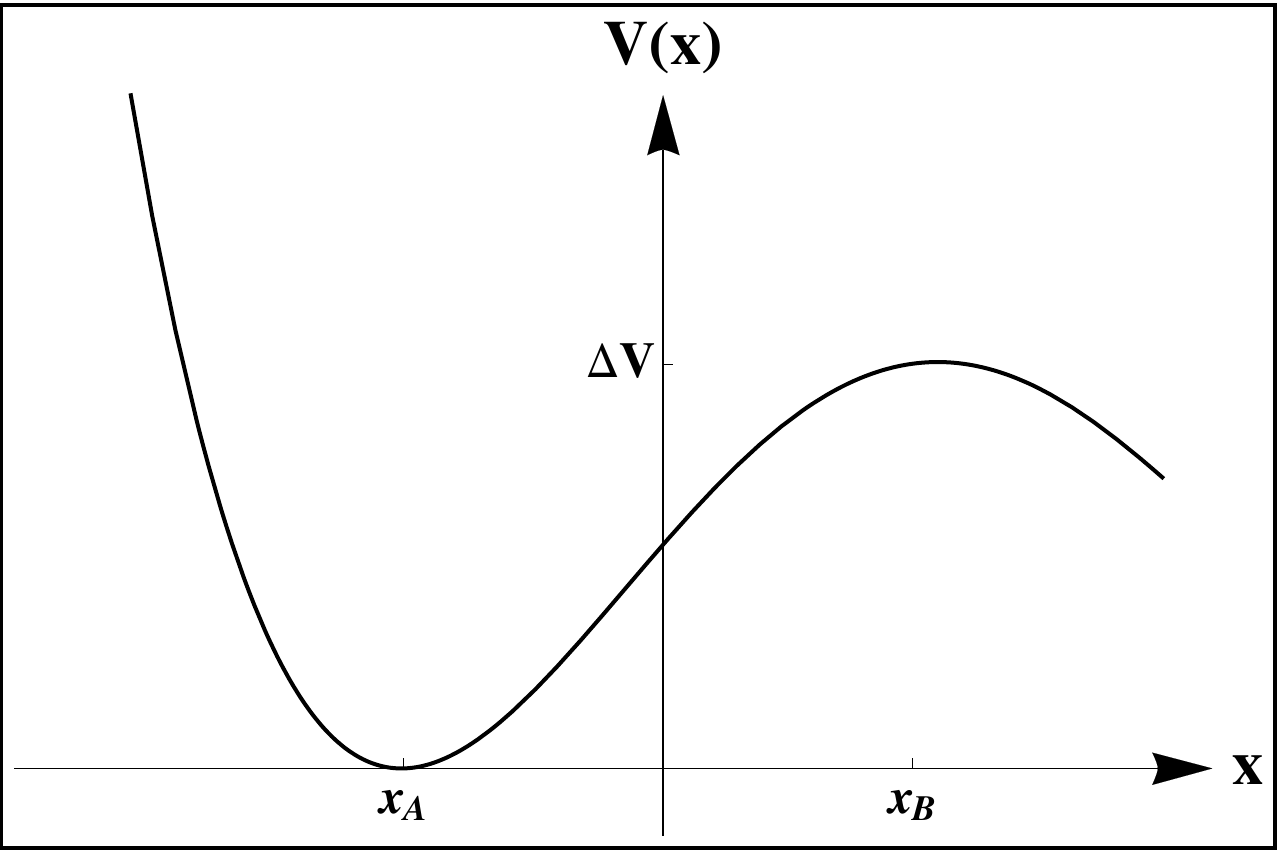}}  
\caption[Potential corresponding to the escape of a classical point
  particle.]{Potential corresponding to the escape problem.  The
  position $x_A$ is the  local minimum, where the point particle sits
  initially, and  $x_B$ the local maximum of the potential.  The
  barrier height is denoted by  $\Delta V$. 
\label{fig:1DPotentialEscape}}
\end{figure}

We consider a classical point particle of mass $m$ initially located
at a local minimum $x_A$ of the potential $V(x)$. {}For simplicity, we
assume  only one direction of escape, which may happen through the
closest local maximum located at $x_B$ on the right of $x_A$. On the
left of  the local minimum, the potential is assumed to rise
and not to have any additional extrema. The situation
is depicted in {}Fig.~\ref{fig:1DPotentialEscape}. Beyond the local
maximum  at $x_B$, to the right the potential might 
have another local or global
minimum or be unbounded from below, it does not matter. 
The height of the barrier is
$\Delta V$.  In particular, the escape rate should be independent of
the shape of the potential beyond the top of the barrier to the
right of $x_B$.

In a classical deterministic description, the particle sitting at the
local minimum stays there forever and an escape from  the potential
well is not allowed. Its dynamics is governed by  Newton's second law
\begin{align}
	m\frac{d^2x}{dt^2} & = -V'(x),
\end{align}
where the prime denotes a derivative with respect to $x$. The position
$x_A$ at the local minimum of the potential is stable. In other words,
$x_A$ is an attractor.  Under a small perturbation, the particle comes
back to the original position.

In the presence of a thermal bath or a fluid, in which the particle is
placed, the situation is altered by the two competing effects
intrinsic  to fluctuation and dissipation dynamics. The random forces,
originating from the thermal fluctuations,  push the particle away
from the initial  position and allow it to eventually climb the
potential barrier.  In addition, the damping tends to slow down the
particle and makes the return to the equilibrium point  $x_A$ more difficult. Due
to the combined effect of fluctuation and  dissipation, the system is
not stable anymore and there is a nonzero probability for the
particle to escape from the well. In particular, after  a sufficiently
long time, it is reasonable to expect that the particle has almost
surely (i.e., it has a nonvanishing probability to have) passed over
the barrier. 
 
We are interested in the rate at which the particle escapes from the
potential well. The escape rate is closely related to the inverse of
the  average time needed to pass, for the first time, the local
maximum of the potential. This time is known in the literature as  the
{``mean first passage time"}~\cite{doi:10.1137/0133024,Talkner1987}. A naive inspection indicates that the escape rate should only depend on the damping coefficient, the strength of the noise, the temperature, and the potential. In regards to the latter, in particular,
the rate depends on the height of the barrier
and the curvature at the minimum and at the maximum. Since the escape is defined from the
first passage at the  top of the barrier, the characteristics of the
potential beyond the maximum should not play any role.
 
{} One clarification on terminology is worth stating here.
For a classical point particle, the escape rate is different from
and should not be confused with a diffusion rate to the next minimum.
The diffusion rate is typically smaller than the escape rate since,
once the particle has passed over the top, it must then go down the
potential on the other side and, eventually, reach the minimum.  If
the next minimum is at  lower energy, the diffusion rate is a decay
rate.  Let us now formulate the
escape problem.

 \subsection{Langevin and Fokker-Planck descriptions}
 
The Langevin and the FP formalisms are two equivalent approaches
used to describe a particle subject to random forces that follow a
Markov process. Both approaches are presented here
with a discussion of their strengths and limitations.

 \subsubsection{The Langevin equation} 
 
The Langevin equation is obtained by the inclusion of the random
force,  parametrized with a  stochastic noise $\xi(t)$, and the
damping term, all in the form of Newton's second law,  
 \begin{align}
 	\label{eq:CH4Langevinequationforthepointparticle}
	m\frac{d^2x}{dt^2} & = - \eta \frac{d x}{dt} -  V'(x) +
        \xi(t),
\end{align}
where  $\eta$  is the damping coefficient. {}For simplicity, the noise
will always be assumed to be Gaussian throughout this work.  The
average over the noise of an operator $\mathcal{O}(x)$ is defined  as
\begin{align}
	\langle\mathcal{O}(x) \rangle_\xi&\equiv\int d [\xi]
        \mathcal{O}(x_\xi)
        \exp\left\{-\frac{1}{2\Omega}\int_{t_0}^{t_f}dt'\xi^2(t')\right\},
\end{align}
where $x_\xi$ is the solution of the Langevin
equation~(\ref{eq:CH4Langevinequationforthepointparticle}) for a given
realization $\xi$ of the noise. Here, $t_0$ and $t_f$ are the initial
and final times.  The noise here satisfies the following relations:
\begin{align}
	\langle \xi(t)\rangle_{\xi}=0, && \langle \xi(t)\xi(t')
        \rangle_{\xi} =\Omega  \delta(t-t'),
\end{align}
where $\Omega$ parametrizes the strength of the noise. The  damping
coefficient $\eta$ is related to $\Omega$ by the Einstein  relation
$\Omega = 2\eta k_B T$. 
 
The Langevin equation is a stochastic differential equation for the
random variable $x_\xi$ and is, therefore, not deterministic.  Given
that the randomness of the stochastic force actually follows a
well-defined probability distribution (in our case, a  Gaussian white
noise Markovian process), the random variable $x_\xi$ should also obey
some probability distribution $\rho(x)$. The FP equation is the
equation whose solution is precisely this probability distribution
$\rho(x)$.  In practice, given that it is a partial differential
equation with well-defined coefficients and boundary conditions, the
FP  equation is better suited for an analytical treatment than the
Langevin equation.
  
 \subsubsection{The Fokker-Planck equation} 
 
 The idea behind the FP description is to consider the evolution of
 the probability distribution of the quantities of interest, in our
 case, the position and the velocity of the particle. Due to the
 presence of random forces, each  realization is achieved with a
 certain probability. Even though each individual particle dynamics
 realization is nondeterministic,  the evolution of the probability
 distribution is deterministic.

We are interested in the position and the velocity of the particle as
a function of time. The Langevin equation gives  a set of two
first-order differential equations for the position $x(t)$ and
velocity $v(t)$,
\begin{align}
	\frac{d x}{dt}  & = v, \\ m\frac{dv}{dt} & = - \eta v - V'(x)
        + \xi(t) .
\end{align}
The FP probability distribution is defined as
\begin{align}
	P(x,v,t \mid x_0,v_0,t_0) &  \equiv \left\langle
        \delta[x_\xi(t)-x]\delta[v_\xi(t)-v]\right\rangle_{\xi} ,
\end{align}
where  the arguments $x_\xi(t)$ and $v_\xi(t)$ of the Dirac
delta-functions on the right are the solutions of the Langevin
equation  (in the presence of the random force $\xi(t)$) and $x$ and
$v$ the arguments of the probability distribution.  $P$ is the
averaged probability to find the particle at position $x$ with
velocity $v$ at time $t$, knowing the initial position  $x_0$ and
velocity $v_0$ at time $t_0$. 

The probability distribution satisfies the FP
equation\footnote{Explicit details on the derivation and a discussion
  about the properties of this equation can be found in
  Refs.~\cite{risken1996fokker,zinn2002quantum}.}
\begin{align}
	\frac{\partial}{\partial t}P(x,v,t \mid x_0,v_0,t_0) & =
        -\mathcal{L}_{FP}P(x,v,t \mid x_0,v_0,t_0)
        , \label{eq:FPEclassicalparticle}
\end{align} 
where $\mathcal{L}_{FP}$ is the FP operator defined as
\begin{align}
	\mathcal{L}_{FP} \equiv  \frac{\partial}{\partial x}v
        -\frac{1}{m}\frac{\partial}{\partial v}\left[\eta v +
          V'(x)\right]  -
        \frac{\Omega}{2m^2}\frac{\partial^2}{\partial v^2}.
\end{align}
The FP equation is an ordinary differential equation for the
probability distribution $P$ and, therefore, analytical methods can be
applied. 

In the large time limit, the system is expected to reach equilibrium.
The equilibrium probability distribution $P_0$ is a time-independent
solution of the FP equation given by 
\begin{align}	
	P_0(x,v) & = \frac{1}{\mathcal{Z}}\exp\left\{-\beta
        \left(\frac{1}{2}mv^2 + V(x)\right)\right\}  \nonumber \\ &=
        \frac{1}{\mathcal{Z}}\exp\left\{-\beta E(x,v)\right\}
        , \label{eq:equdistcp}
\end{align}
 where $E$ is the energy of the nondissipative system and the
 partition function $\mathcal{Z}$ is the normalization.  Note that the
 equilibrium distribution always formally exists as a solution of the
 FP equation, however, it does not  necessarily  imply that the system
 possesses an equilibrium state. The equilibrium distribution can be
 non-normalizable,  in particular, if the potential is unbounded from
 below.  The FP formalism  is fully equivalent to the Langevin
 approach and provides the tools needed for an analytical  derivation
 of the escape rate.

 \subsection{Computation of the escape rate}
 
 Over the last century, several methods have been proposed to estimate
 the escape rate.\footnote{For a comprehensive review  of these
   methods, see, e.g., Ref.~\cite{RevModPhys.62.251}.} Since our
 final goal is to consider a  relativistic scalar field, we focus on
 the flux-over-population method that appears as the most promising
 candidate for  such a generalization. {}For a better interpretation
 of the escape problem, we introduce, in the Appendix, the framework of the
 MFPT and show its formal equivalence with
 the flux-over-population method, which proves that the escape  rate is
 indeed the inverse of the MFPT.

 \subsubsection{Flux-over-population method}
 \label{sec:fopmonedim}
 
The flux-over-population method has been introduced in
Ref.~\cite{L.:1927aa} and the explicit computation of the rate  has
been achieved by Kramers in Ref.~\cite{KRAMERS1940284}.

\begin{figure}[htb!]\centering
 {\includegraphics[width=0.75\columnwidth]{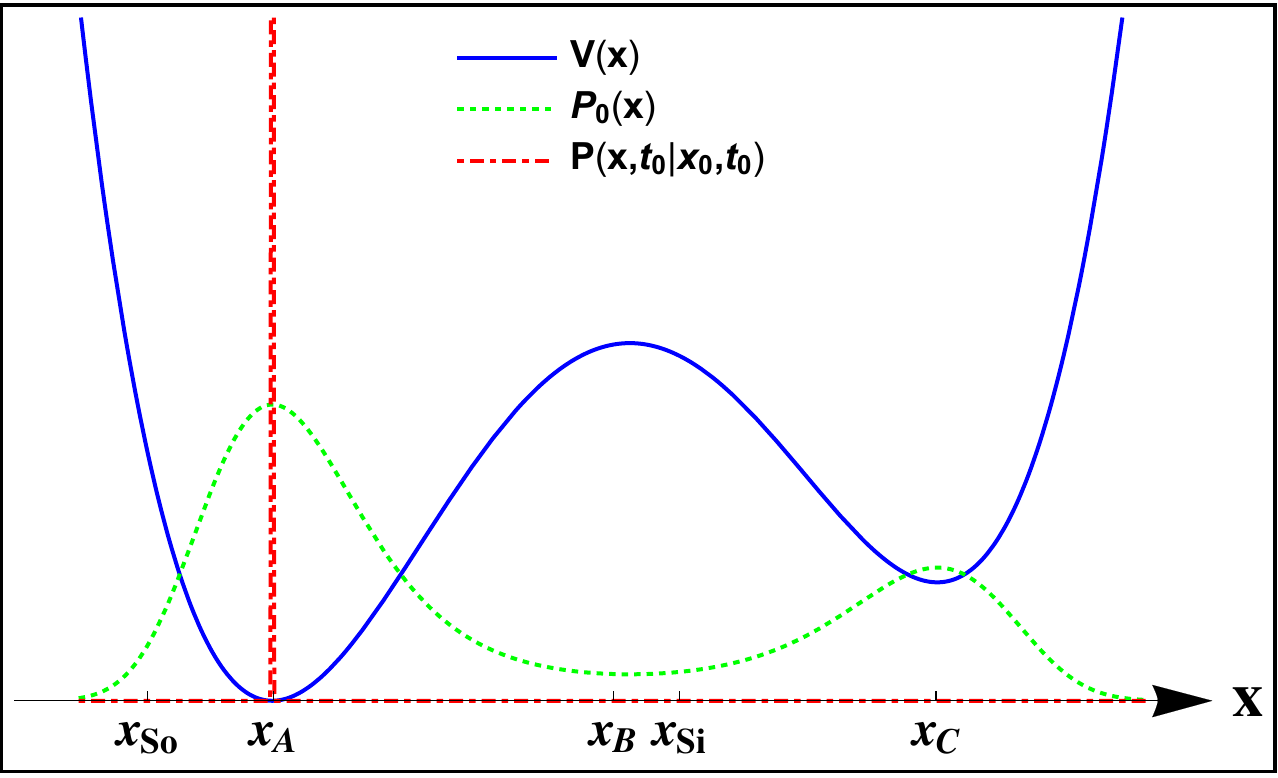}}  
\caption[Flux-over-population method.]{Example case studied
  with the flux-over-population method.  The blue solid line is the
  potential $V(x)$. The red dashed-dotted line is the particle's initial 
position and the
  green dotted  line is the equilibrium FP probability distribution
  for the position. The position $x_A$ is the  initial location of the
  particle, $x_B$ the local maximum, $x_C$ a second local minimum,
  $x_{So}$ and $x_{Si}$  the positions of the source and the sink
  respectively. \label{fig:1DFOPM}}
\end{figure}

Let us consider the situation shown in {}Fig.~\ref{fig:1DFOPM}. {}For
illustrative purposes, we have chosen  an asymmetric double-well
potential. Similar reasoning applies to any kind of potential as long
as it possesses  a local minimum in the vicinity of a local
maximum.\footnote{The shape of the potential influences the form of
  the equilibrium distribution, however, the existence of a
  probability flux at the top of the potential is  guaranteed.}  The
particle is initially located at the minimum $x_A$ and the  FP
probability  distribution at time $t_0$ is a product of two Dirac
delta-functions,
\begin{align}
	P(x,v,t_0 \mid x_A,0,t_0) & = \delta(x-x_A)\delta(v).
\end{align}
The position-dependent parts of the initial and equilibrium
probability distributions  are shown in {}Fig.~\ref{fig:1DFOPM},  with
the red dashed-dotted line and the green dotted line, respectively.
During the time evolution, given by the FP
equation~\eqref{eq:FPEclassicalparticle},  the probability
distribution goes from the red dashed-dotted line to the green dotted line.  In the large
time limit, the system has reached  equilibrium and the probability
distribution is given by  Eq.~\eqref{eq:equdistcp}.  Therefore, there
must be a flux of probability at the maximum of the well. The origin
of this flux of probability is precisely  the fluctuation and
dissipation dynamics discussed previously. 

The idea behind the flux-over-population method relies on the
construction of a steady-state 
solution. The inclusion of sources and sinks maintains a constant probability current across the well. The role
of the sources, located  to the left of the minimum at $x_{So}$
in {}Fig.~\ref{fig:1DFOPM}, is to supply particles 
to the ``$A$-well'' and maintain
a constant number density inside the well. The particles
thermalize and eventually leave the well before being removed by the
sinks, located  on the right of the maximum at $x_{Si}$. Since the
total probability flux $j$ is equal to the rate of escape $k$  times
the population of the $A$-well, $n_A$, the flux-over-population method
predicts
\begin{align}
	\label{eq:fop1dformula}
	k \equiv  \frac{j}{n_A},
\end{align}
as a solution for the escape rate. 

The population of the $A$-well is given by the integration over the
probability density,
\begin{align}
	n_A & = \int_{A-\text{well}} dx dv\ P(x,v),
\end{align}
which corresponds to the probability to be in the well, with
$x\in(-\infty,x_B]$ and $v\in (-\infty, +\infty)$. The flux at the
        barrier is 
\begin{align}
	j & = \int_{-\infty}^{+\infty} dv\ v P(x_B, v),
\end{align}
which is the probability to pass over the maximum with some velocity. 

The derivation of the rate requires two steps. First, the
probability distribution  is obtained and then second the flux and the  number
density are computed. The probability density $P$ is a solution of the FP
equation~\eqref{eq:FPEclassicalparticle} with the  particular boundary
conditions dictated by the specific steady-state situation under
consideration. The ensemble of particles is  in equilibrium inside the
$A$-well and the probability density is given by
Eq.~\eqref{eq:equdistcp}. Since the sinks remove the  particles once
they have passed the maximum, we impose 
		\begin{align}\label{eq:absorbing-BC}
			P( x > x_{Si},v) & \simeq 0.
		\end{align}
{}Finally, at the top of the barrier, there are no sources or sinks and
the potential $V(x)$ is approximated as
		\begin{align}
			V(x)\simeq V(x_B) - \frac{1}{2}
                        |V''(x_B)|(x-x_B)^2+\mathcal{O}[(x-x_B)^3],
		\end{align}
		 and, therefore, the steady-state FP
                equation~\eqref{eq:FPEclassicalparticle} becomes
		\begin{align}
	&\left\{- \frac{\partial}{\partial x}v  +
                  \frac{1}{m}\frac{\partial}{\partial v}\left[\eta v
                    -|V''(x_B)|(x-x_B)\right]  \right.  \nonumber \\ &
                  \left. +
                  \frac{\Omega}{2m^2}\frac{\partial^2}{\partial
                    v^2}\right\}P(x,v)  = 0,
		\end{align}
 at the vicinity of the top of the barrier $x_B$.

The construction of $P(x,v)$ relies on the Kramers
ansatz~\cite{KRAMERS1940284},
\begin{align}
	P(x,v) & = \zeta(x,v) P_0(x,v),	
	\label{eq:defzetaxv}
\end{align}
where $P_0$ is the equilibrium distribution and $\zeta$ is chosen to
satisfy the boundary conditions:
\begin{align}
	\lim_{x\rightarrow x_A}  \zeta(x,v) & = 1, &&
 	\zeta(x>x_{Si},v)  = 0. 
	  \label{eq:FPKBC}
\end{align}
Applying the FP operator on the ansatz and using the equilibrium
distribution Eq.~\eqref{eq:equdistcp}, we obtain the equation for
$\zeta$,
\begin{align}
	& \left\{- v  \frac{\partial }{\partial x} -
  \frac{1}{m}\left[\eta v + |V''(x_B)|(x-x_B)\right]
  \frac{\partial}{\partial v}  \right.  \nonumber \\ & \left. +
  \frac{\Omega}{2m^2}\frac{\partial^2}{\partial v^2} \right\}
  \zeta(x,v)  = 0 ,
  \label{eq:FPKansatz}
\end{align}
where we identify the adjoint FP equation 
\begin{align}
	 \mathcal{L}_{FP}^\dagger\zeta(x,v)  = 0.
\end{align} 
 In order to  solve
this equation, Kramers made the further assumption in~\cite{KRAMERS1940284} that $\zeta$
depends only on $u$, a linear combination of $x$ and $v$, such that
\begin{align}
	u & \equiv (x-x_B) + a v .
	\label{eq:Kramersassumption}
\end{align}
From a purely mathematical point of view, this assumption allows finding a solution of the differential equation~\eqref{eq:FPKansatz} that satisfies the boundary conditions~\eqref{eq:FPKBC}. According to Ref.~\cite{LANGER1969258}, it can be shown that the form of solution $\zeta(u)$ is unique. 

A physical interpretation for adopting the ansatz~\eqref{eq:FPKansatz} can be obtained by looking at the function $\zeta(x,v)$. From its definition~\eqref{eq:defzetaxv}, $\zeta$ parametrizes the deviation from equilibrium due to thermal activation in the vicinity of the saddle point. The second boundary condition in~\eqref{eq:FPKBC} implies that  $\zeta(x,v)$ should go to zero in the region of phase space away from the saddle point (the "probability sink" at $x\gg x_B$) and also should quickly vanish in the region of positive velocities. Away from the saddle point, it is fair to expect that the vanishing of the function $\zeta(x,v)$ is controlled by either $(x-x_B)$ or $v$. The linear combination in the Kramers assumption~\eqref{eq:Kramersassumption} is the simplest and most straightforward way to implement this idea. The equation for $\zeta(u)$ then becomes
\begin{align}
	-\left[ (1 + \frac{a}{m}\eta )v +  \frac{a}{m}
          |V''(x_B)|(x-x_B)  \right] \zeta'+  a^2 \frac{\Omega}{2m^2}
        \zeta'' & = 0 ,
\end{align}
where the prime denotes a $u$-derivative. {}For consistency with the
assumption that  $\zeta$ is a function of $u$ only and in order to
obtain the correct  behavior at the boundary, the factor in front of
$\zeta'$ must be a linear function of  $u$. Imposing that
\begin{align}
	\lambda u & \equiv-\left[ (1 + \frac{a}{m}\eta )v +
          \frac{a}{m} |V''(x_B)|(x-x_B)  \right]
        , \label{eq:definitionlambdaucp}
\end{align}
 the constants  $a$ and $\lambda$  are found to be given,
 respectively, by 
\begin{eqnarray}
\label{eq:lambdappc}
&&	\lambda_{\pm}  =
  -\frac{\eta}{2m}\pm\sqrt{\frac{|V''(x_B)|}{m}+\left(\frac{\eta}{2m}\right)^2},
  \\ && a = -\frac{m}{V''(x_B)} \lambda_\pm,
\end{eqnarray}
where the two solutions for $\lambda$ have opposite signs.

Solving for $\zeta(u)$ by inserting Eq.~\eqref{eq:definitionlambdaucp}
in the differential equation and integrating twice gives
\begin{align}
	\zeta(u) & = \sqrt{\frac{\beta[V''(x_B)]^2}{2 \pi \eta
            \lambda_+}} \int_{u}^{\infty} dz\ \exp\left\{-\beta
        \frac{[V''(x_B)]^2}{2\eta \lambda_+}z^2\right\},
\end{align}
where $\lambda_+$ has been chosen to have an overall negative exponent
and, therefore, $\zeta$ to vanish for large positive $x$. The  factor
in front of $\zeta$ has been chosen to satisfy the condition that
$\zeta$ goes to unity inside the $A$-well.

Having all elements at our disposal to compute the probability flux $j$,
we obtain the result
\begin{align}
	j & =    \frac{1}{\mathcal{Z}} \left(\frac{
          \lambda_+}{\beta}\right)\frac{1}{\sqrt{m |
            V''(x_B)|}}\exp\left\{-\beta V(x_B)\right\} ,
\end{align} 
where we have used integration by parts. The population $n_A$ of the
$A$-well is simply
\begin{align}
	n_A &    \simeq \frac{1}{\mathcal{Z}}\sqrt{\frac{2\pi}{\beta
            m}}\sqrt{\frac{2\pi}{\beta V''(x_A)}}\exp\left\{-\beta
        V(x_A)\right\},
\end{align}
where the potential has been expanded around the local minimum in
$x_A$ and  the limit of integration for $x$ safely extended to
infinity.

Taking the ratio of $j$  and $n_A$, the escape rate is found to be
\begin{align}
	\label{eq:CH4fopratepointparticlefinal}
	k &  =
        \frac{\sqrt{\frac{|V''(x_B)|}{m}+\left(\frac{\eta}{2m}\right)^2}-
          \frac{\eta}{2m}}{2\pi}
        \sqrt{\frac{V''(x_A)}{|V''(x_B)|}} \nonumber \\ & \times
        \exp\left\{-\beta \left[V(x_B)-V(x_A)\right]\right\},
\end{align}
which is the famous result of Kramers. As expected, the rate depends
only on the parameters $\eta$ (or equivalently $\Omega$), the
temperature, the curvature of the potential at the initial  local
minimum and the nearby maximum and the height of the barrier.  The
height of the barrier $\Delta V = V(x_B)-V(x_A)$ can be seen as the
activation energy. 

It is important  to notice that the shape of the potential on the
other side of the well does not influence the final result. This is a
consequence of the absorbing boundary condition
(\ref{eq:absorbing-BC}). Notice that this feature is crucial for a
sound definition of the escape problem.  Indeed, for a potential well
such as the one shown in {}Fig.~\ref{fig:1DFOPM}, there is a minimum
at $x_C$, which has a higher potential energy compared to the starting
well at $x_A$. In this case, the total probability flow over the
barrier at $x_B$ will be made up of a sum of two contributions: one
escaping contribution (from $x_A$ to $x_C$) and a returning
contribution (from $x_C$ back to $x_A$). The absorbing boundary
condition (\ref{eq:absorbing-BC}) is a way to disentangle the two
contributions, keeping only the escaping part of the probability flux
over the barrier. 

An alternative method to derive the escape rate is achieved with the formalism of the mean first passage time. This approach provides a simple interpretation of the escape problem; it corresponds to the average time needed to leave for the first time a specified domain. In practice, it is difficult to solve for the MFPT; this is due, in particular, to nontrivial boundary conditions. In the Appendix, we provide more details on the formalism of the MFPT and show the formal equivalence with the flux-over-population method.


 \section{Escape Rate for a Scalar Field}
 \label{sec:escaperateforscalarfieldderivation}
 
The main objective of this work is the definition of the Kramers
problem in field theory.  Using the knowledge gathered from the
classical point particle case, we first describe the escape problem
for a scalar field and  then show that the formulation of a meaningful
definition is not straightforward. 

We introduce the two usual formulations for dealing with the Kramers
problem. The first is based on the Langevin equation, which
has a direct interpretation but is limited in its analytical
treatment. The second formulation is the one based on the FP
approach, whose derivation is more involved, but is
much more amenable to analytic treatment.
In this section we will also use ideas from the
flux-over-population method to define the Kramers problem, derive
explicitly the escape rate for a scalar field and then interpret  the
results.

\subsection{Defining the escape problem for a background scalar
  field configuration}
 \label{sec:CHdefinitionoftheescapaeratefield}

\begin{figure}[htb!]\centering
 {\includegraphics[width=0.75\columnwidth]{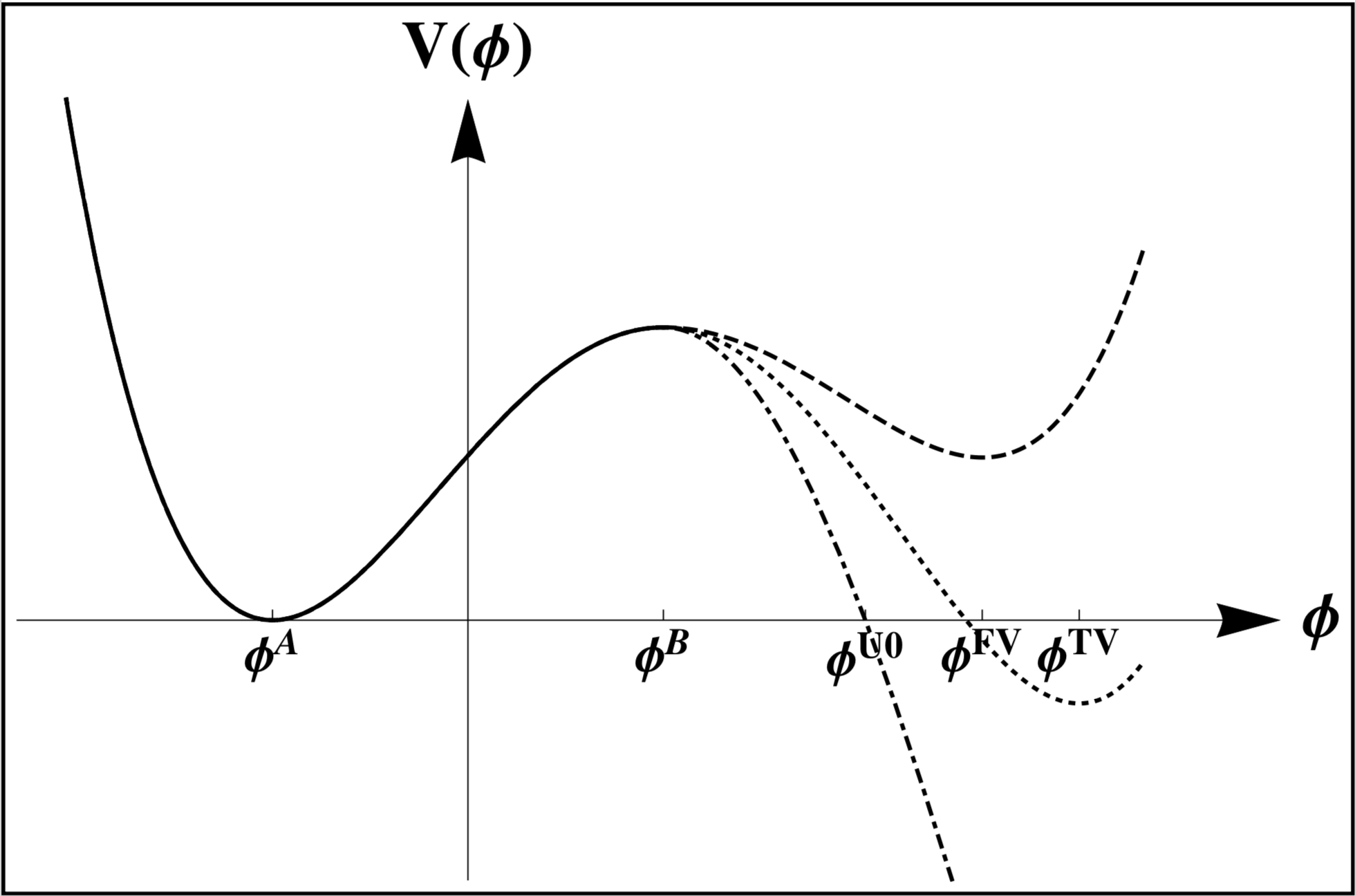}}  
\caption[Potentials corresponding to the escape problem of a
  field.]{Potentials corresponding to the escape problem. $\phi^A$ is
  the initial local minimum.  $\phi^B$ corresponds to the local
  maximum of the potential. On the right-hand side of the maximum,
  there are several possibilities, a false vacuum at $\phi^{FV}$
  (dashed line), a true vacuum at  $\phi^{TV}$ (dotted line) or a potential that is 
  unbounded from below (dashed-dotted line) with $V(\phi^{U0}) =
  V(\phi^{A})$.\label{fig:ScalarPotentialEscape}}
\end{figure}

Let us consider a scalar field with a self-interaction potential as
shown in {}Fig.~\ref{fig:ScalarPotentialEscape}.  We assume, for
simplicity, that the initial configuration is a homogeneous field
sitting at a local minimum $\phi^A$.  The interactions with extra
degrees of freedom, for example a thermal bath, lead to fluctuation
and dissipation  dynamics and potentially allow for an escape from the
potential well. Our goal is to compute the rate per unit  volume for
the field to escape from the  well, due to thermal fluctuations. 

Involving a field renders the definition of an escape more difficult
and less intuitive than treating the point particle, zero-dimensional case,
discussed previously. At equilibrium,
the field populates both sides of the well (or has completely decayed
if the potential  is unbounded from below beyond the well). Comparing with the initial situation, where the field configuration is homogeneously located at $\phi^A$, it is
reasonable to assume the existence of a flow of the probability
density  across the potential well. For this reason, the flux-over-population method should apply. The naive generalization of the
point particle case would be to estimate the average time needed for
the field to reach the top of the potential $\phi^B$  for the first
time at each point in space. As we will learn, this case can be
related to the Hawking-Moss  solution~\cite{HAWKING198235} in the
early Universe. However, in our case, where we are considering a
Minkowski spacetime and the volume  can be infinite, this solution
would lead to a vanishing rate. We should, therefore, seek for another
definition of the  escape configuration.

Before going into the details of our calculations, it is important to
comment on the difference between the escape problem treated in this
work with two other closely related problems, the quantum tunneling
and the nucleation problem. Quantum tunneling, as its name implies, is
a consequence of the quantum fluctuations of the field. Such
fluctuations can connect two classically disconnected  values of the
field, through a forbidden region in potential energy, without giving
the field any energy.  This is what happens in a quantum first-order
phase transition. Nucleation, on the other hand, is the  mechanism
that drives first-order phase transitions with small degrees of
metastability (for example,  small supercooling). It corresponds to
the formation (or ``nucleation'') of bubbles of the stable phase
inside the metastable phase. Such bubbles grow and complete the phase
conversion. Differently from  tunneling, the process of nucleation is
typically driven by thermal fluctuations (even though for many systems
quantum fluctuations may also play a role). In this sense, it can be
said that, in nucleation,  the potential energy barrier is overcome
with energy absorbed from the heat reservoir, in  contrast to
tunneling. Lastly, the problem treated in this work, the  escape
problem,   does not necessarily require the presence of an initial false vacuum. If it is the case, the escape problem 
can be seen as the first stage of the nucleation problem,
i.e., the generation of  domains of field configurations outside the
initial minimum. In general, the Kramers problem for a scalar field, defined in this work, should be understood as the derivation of the probability for the field to pass over the potential barrier  in a finite region of space. Due to thermal fluctuations (quantum fluctuations
could also be eventually be considered in a future work), a  field
starting at a low minimum of potential energy can gain energy from the
heat reservoir and then ``climb'' the potential well to reach and
surmount an energy barrier. Another distinctive feature
of the escape problem is that it does not regard the dynamics of the
field after it finds itself outside of the  starting well. However,
this issue is more subtle for a field than for a point particle
system,  as we will discuss below.

As stated in Ref.~\cite{zinn2002quantum}, it is sufficient that a
finite part of space has passed the barrier. At first  sight, this
statement would give some freedom in the precise definition of the
escape problem. In particular, once the  field has reached the top at
a spatial location, it can fall on the other side and attract the
neighboring points,  with a gain in potential energy, likely accompanied by an energy cost due to inhomogeneities, i.e., a nonzero gradient term. This is a
crucial difference with the zero-dimensional, one-particle case of the
previous section. When considering a field, the form of the potential
beyond the maximum plays a role  in the escape problem. Indeed, the
Laplacian term in the field equation of motion connects neighboring
points of space, which tend to have close values of the field and are
thus not independent of  each other. Therefore, if a given point of
space acquires a field value beyond the maximum of the potential, this
is in a way of saying it ``communicated'' to its neighboring
points. This is a distinctive trait of the field system's dynamics that
differs  from the
single particle case.  It is then fair to expect that
the two cases, where the initial minimum is a true or a false vacuum,
must be treated  separately.  As we will learn shortly, these features
naturally emerge along the computation in a generalized
flux-over-population method and this approach allows for a
satisfactory definition of the escape problem. In  particular, a
critical volume of space that experiences hopping is precisely defined
by the formalism.  To perform  this analysis, we should first
introduce some of the necessary ingredients from stochastic field
theory. 

 \subsection{Stochastic field theory}
 \label{sec:secstocasticFT}
 
Let us review here some of the basics of stochastic field theory and
introduce the relevant quantities needed for the derivation of the escape rate.
  
  \subsubsection{The Langevin and Fokker-Planck equations}

The dynamics of a classical field configuration under random
fluctuations and dissipation is an extremely important subject in many
different branches in modern physics as far as the description of
nonequilibrium fields is concerned (for a thorough introduction and
review, see, e.g., Ref.~\cite{calzetta+hu}).   A natural
characteristic when studying the evolution of a system in interaction
with an environment is the presence of both dissipative and stochastic
terms. {}For instance, in the context of quantum fields, we might be
interested in the derivation of an effective equation of motion for a
given field background configuration which represents some relevant
characteristic of the system under study (e.g., a vacuum expectation
value taking the role of an order parameter important in a phase
transition problem). Typically, this involves a selection of a
relevant field mode, in which we are interested in the dynamics and
will represent the physical system, while the remaining degrees of
freedom are taken as an environment.  The degrees of freedom that
are regarded as environment can also include any other fields in the
original model Lagrangian (see for instance Ref.~\cite{calzetta+hu}
for a review and discussion about these types of equations
and their derivation in the context of quantum field theory). In
quantum field theory, the preferred methodology used to study
dynamical effects in general is the closed time path
formalism~\cite{Chou:1984es}.  The effective equation of motion for an
interacting scalar field is Langevin-like, and includes an
explicit fluctuation-dissipation relation (see, e.g.,
Refs.~\cite{Morikawa:1986rp,hu1,Gleiser:1993ea,Berera:1998gx,Berera:2007qm}).
Generically, the usual relativistic Klein-Gordon equation describing
the dynamics of the scalar field in a potential $V(\phi)$ is modified
to take the  thermal fluctuations into account and becomes a Langevin
equation,
 \begin{align}
 	\label{eq:LE}
 	(\partial_t^2-\nabla^2)\phi(\vec{x},t)+\frac{\partial
          V(\phi)}{\partial\phi}+\eta
        \dot{\phi}(\vec{x},t)&=\xi(\vec{x},t),
 \end{align}
 where  $\eta$ is the dissipation coefficient and $\xi$ is a
 Gaussian white noise satisfying 
 \begin{align}
 	&\langle \xi(\vec{x},t) \rangle = 0, \nonumber \\ &\langle
   \xi(\vec{x},t) \xi(\vec{x}',t')\rangle = \Omega
   \delta^3(\vec{x}-\vec{x}')\delta(t-t'),
 \end{align}
 where $\Omega$ parametrizes the strength of the noise and satisfies
 the Einstein relation $\Omega=2\eta /\beta=2\eta k_BT$.  In the
 quantum field theory context, both the potential and the dissipation
 coefficient in Eq.~(\ref{eq:LE}) can be functions of the temperature
 and on the details of the interactions in the full original
 microscopic Lagrangian density, carrying, for example, information on
 the interactions of the scalar field $\phi$ with other field degrees
 of freedom.  In the following, we will assume a particular fixed form for the
 potential and the dissipation coefficient.  It is
straightforward to generalize the analysis
for other forms, for example, that include the
 dependence on the temperature.  Exploring the full
 quantum origin of the Langevin equation for the expectation value of
 a field goes beyond the scope of this work. We refer the interested
 reader to ~\cite{calzetta+hu} and references
 therein. Thus, for the rest of this work, we will simply assume the
 existence of a Langevin equation of the form of Eq~\eqref{eq:LE}.
 
{}As an important point of detail, when dealing with integrals over the
field space, we discretize the space by adopting the following
conventions, 
  \begin{align}
 	&\phi(\vec{x},t) \rightarrow \phi({x}_i,t) = \phi_i(t),
    \nonumber \\ &\int d^3\vec{x} \rightarrow a^3 \sum_{i=1}^{N^3},
    \nonumber\\ & \delta(\vec{x}-\vec{y}) \rightarrow
    \frac{\delta_{ij}}{a^3}, 
	\end{align}
such that 
\begin{align}
 	&\int d^3\vec{x}\ \delta(\vec{x}-\vec{y})   =  a^3
  \sum_{i=1}^{N^3} \frac{\delta_{ij}}{a^3} =1,
 \end{align}
 where the volume $\mathcal{V} = L^3 = ( N\cdot a )^3$, with $N$
being the number of discrete sites in each direction and $a$  the
spacing between two adjacent points.  Taking into account the field and its conjugate momentum, in a space of three dimensions, we are considering a $2N^3$-dimensional system. {}For simplicity, we have labeled
the spatial points in the three spatial directions with a  single
label $_i$ instead of $_{ijk}$. {}For the sake of clarity, we will
also denote the Laplacian as $\nabla_{ij}^2\phi_j$. The  actual
definition in discrete space is given by 
  \begin{align}
 	\nabla^2\phi_{ijk} & =
        \frac{1}{a^2}\left[\phi_{i+1,j,k}+\phi_{i-1,j,k}+\phi_{i,j+1,k}
          \right.  \nonumber \\ &
          \left. +\phi_{i,j-1,k}+\phi_{i,j,k+1}+\phi_{i,j,k-1}-
          6\phi_{i,j,k}\right],
 \end{align}
 where each direction of space has been explicitly labeled.
 
As usual when working with the Langevin equation like in
Eq.~(\ref{eq:LE}), the analytical treatment is limited by the
stochastic character of  the equation. There is, therefore, a need to
introduce the FP formalism for the scalar field.  The Langevin
equation (\ref{eq:LE}) implies the following set of equations for the
field $\phi$ and the conjugate momentum $\pi$,
 \begin{align}
 	\label{eq:LEE}
 	\partial_t\phi_i(t)&=\pi_i(t),\notag\\ \partial_t\pi_i(t)&=
        -\eta \pi_i(t)+\nabla_{ij}^2\phi_j(t)-V'(\phi_i)+\xi_i(t),
 \end{align}
 where the prime denotes a derivative with respect to the field. The
 FP probability density is defined as
  \begin{align}
 	P(\phi,\pi,t\mid\phi_0,\pi_0,t_0)&\equiv\left\langle
        \prod_{i=1}^{N^3}\delta
             [\hat{\pi}_i(t)-\pi_i]\cdot\delta[\hat{\phi}_i(t)-\phi_i]\right\rangle_\xi,
 \end{align}	
where $\hat{\phi}_i(t)$ and $\hat{\pi}_i(t)$ are solutions of the
Langevin equation~\eqref{eq:LEE} for a given noise realization $\xi$
and $\phi_i$ and $\pi_i$ are the arguments of the probability
distribution $P$. The stochastic  average of an operator
$\mathcal{O}(\hat{\phi},\hat{\pi})$ is defined as 
\begin{align}
	\left\langle
        \mathcal{O}(\hat{\phi},\hat{\pi})\right\rangle_\xi\equiv &
        \int \prod_{i=1}^{N^3}d[\xi(t)]_i
        \mathcal{O}(\hat{\phi},\hat{\pi})  \nonumber \\ & \times
        \exp\left\{ -\frac{a^3}{2\Omega} \sum_{j=1}^{N^3}\int dt'
        \xi_j^2(t')\right\}, \label{eq:noiseaverage}
\end{align}
where the integration measure is normalized to give $\langle 1
\rangle_\xi =1$. The probability density is a solution of the FP
equation,
   \begin{align}
  	\label{eq:FPEscalarfield}
 	\frac{\partial }{\partial t} P(\phi,\pi, t\mid\phi_0,\pi_0,
        t_0) & =-\mathcal{L}_{FP} P(\phi,\pi, t\mid\phi_0,\pi_0, t_0)
        ,
 \end{align}
 where the FP operator is defined as
    \begin{align}
 \mathcal{L}_{FP} \equiv & - a^3\sum_{i=1}^{N^3} \left\{-\pi_i
 \frac{\partial}{a^3\partial \phi_i} \right.  \nonumber \\ &
 \left. +\frac{\partial}{a^3\partial \pi_i}\left[\eta \pi_i -
   \nabla_{ij}^2\phi_j +V'(\phi_i)\right]+\frac{\Omega}{2}
 \frac{\partial^2}{ a^6\partial\pi_i^2}\right\}.
  \end{align}
  
\subsubsection{The probability density current} 

Due to the conservation of probability, the FP equation can be written
in terms of a probability density current $J$,
\begin{align}
	\label{eq:FPSFFPcurrent}
	\partial_t P(\phi, \pi,t) & = - a^3 \sum_{i=1}^{N^3}
        \frac{\partial}{a^3 \partial \phi_i} J_i  - a^3
        \sum_{i=1}^{N^3} \frac{\partial}{a^3 \partial \pi_i}
        \bar{J}_{i},
\end{align}
where the components  $J_i$ and $\bar{J}_i$ of the current are defined
as 
\begin{align}
	J_i  & \equiv -  \left\{-\pi_i -  k_B T
        \frac{\partial}{a^3\partial \pi_i}\right\} P(\phi,
        \pi,t\mid\phi_0, \pi_0,t) ,\label{eq:ProbabilityCurrentJ}
	\end{align}
and
\begin{align}
	\bar{J}_i  \equiv & - \left\{ \left[  \eta \pi_i-
          \nabla_{ij}^2\phi_j +V'(\phi_i) \right] + k_B T
        \frac{\partial}{a^3\partial \phi_i } \right.  \nonumber \\ &
        \left. +\frac{\Omega}{2}  \frac{\partial}{a^3\partial \pi_i}
        \right\} P(\phi, \pi,t\mid\phi_0, \pi_0,t)
        , \label{eq:ProbabilityCurrentJbar}
\end{align}
for $i\in[1,N^3]$. The validity of this equation can be shown
explicitly by substituting in Eq.~\eqref{eq:FPSFFPcurrent}.

\subsubsection{The equilibrium distribution} 

The FP equation admits an equilibrium solution $P_0$ given by
\begin{align}
	\label{eq:equilibriumdistributionfield}
	P_0(\phi,\pi) & = \mathcal{Z} ^{-1} \exp\left\{-\beta
        E[\phi,\pi]\right\},
\end{align}
where $\mathcal{Z} $ is the normalization given by the partition
function
\begin{align}
	\mathcal{Z} & = \int \prod_{i=1}^{N^3} d\phi_i d\pi_i
        \exp\left\{-\beta E[\phi,\pi]\right\},
\end{align}
and $ E[\phi,\pi]$ is
\begin{align}
		 E[\phi,\pi] & = a^3 \sum_{i=1}^{N^3}\left[
                   \frac{1}{2} \pi_i^2 + \frac{1}{2} (\nabla \phi_i)^2
                   + V(\phi_i)\right],
\end{align}
which corresponds to the energy function of the system in the
nondissipative limit. 

\subsubsection{The vector-matrix notation} 

{}Following the work of Langer~\cite{LANGER1969258}, it is useful to
introduce a vector-matrix notation. The field  and its conjugate
momentum are written in a $2N^3$-dimensional vector as
\begin{align}
	\left(  \begin{array}{c} \phi \\  \pi \end{array}\right) =
        \left(  \begin{array}{c} \phi_i(t)
          \\  \pi_i(t)\end{array}\right) , && \text{where }
        i\in[1,N^3].
\end{align}
The deterministic limit of the Langevin equation is expressed as
\begin{align}
	\frac{\partial}{\partial t}\left(  \begin{array}{c} \phi
          \\  \pi \end{array}\right) & =
        -M\cdot\left(  \begin{array}{c}
          \frac{\partial}{a^3\partial\phi}
          \\  \frac{\partial}{a^3\partial\pi} \end{array}\right)
        E[\phi,\pi],
\end{align}
  with $M=(M_{ij})$ being the $2N^3\times 2N^3$ block matrix defined
  as
  \begin{align}
	M & = \frac{1}{a^3}\left(  \begin{array}{cc} 0 & -\mathbb{1}
          \\ \mathbb{1} & \eta \mathbb{1} \end{array}\right),
\end{align}
where $\mathbb{1} $ is the $N^3$-dimensional unit matrix and the
multiplication $\cdot$  between two $2N^3\times 2N^3$ matrices is
defined as 
\begin{align}
	\label{eq:scalarproductmatrixfield}
	(A\cdot B)_{ij} & \equiv a^3 \sum_{k=1}^{2N^3} A_{ik} B_{kj}.
\end{align}
 A similar rule applies to the scalar product. The FP equation is
 given as
\begin{align}
	\label{eq:FPSFFPcurrentmatrix}
	\partial_t P(\phi, \pi,t) & = - \left(  \begin{array}{c}
          \frac{1}{a^3} \frac{\partial}{\partial\phi}
          \\  \frac{1}{a^3}
          \frac{\partial}{\partial\pi} \end{array}\right)^T \cdot
        \left(  \begin{array}{c} J \\  \bar{J} \end{array}\right)  ,
\end{align}
where $(J\ \bar{J})^T$ is the $2N^3$-dimensional vector corresponding
to the probability current
\begin{align}
	 \left(  \begin{array}{c} J \\  \bar{J} \end{array}\right)  &
         = - M \cdot \left(  \begin{array}{c} \frac{1}{a^3}
           \frac{\partial E}{\partial\phi} + \frac{k_BT}{a^3}
           \frac{\partial }{\partial\phi} \\  \frac{1}{a^3}
           \frac{\partial E}{\partial\pi}  + \frac{k_BT}{a^3}
           \frac{\partial}{\partial\pi} \end{array}\right) P(\phi,
         \pi,t\mid\phi_0, \pi_0,t).
\end{align}

 \subsubsection{The continuum limit}  

We have been working in discrete space to simplify the analytical
computations. However, the continuum limit can be taken at any stage
of the derivation. {}For completeness, let us state the main
quantities as expressed in the continuum limit. The FP equation
reads
    \begin{align}
 	\frac{\partial }{\partial t}  P(\phi,\pi, t\mid\phi_0,\pi_0,
        t_0)  & = -\mathcal{L}_{FP}P(\phi,\pi, t\mid\phi_0,\pi_0,
        t_0),
 	\end{align}
 	with
 	\begin{align}
	\mathcal{L}_{FP} &\equiv -   \int d^3\vec{x}
        \left\{-\pi(\vec{x}) \frac{\delta}{\delta\phi(\vec{x})}
        \right.  \nonumber \\ & \left. +\frac{\delta}{\delta
          \pi(\vec{x})}\left[\eta \pi(\vec{x}) - \nabla^2\phi(\vec{x})
          +V'(\phi)\right]+\frac{\Omega}{2} \frac{\delta^2}{
          \delta\pi(\vec{x})^2}\right\} ,
  \end{align}
 and the equilibrium distribution is given by 
\begin{align}
	P_0(\phi,\pi) & = \mathcal{Z} ^{-1} \exp\left\{-\beta
        E[\phi,\pi]\right\},
	\end{align}
	with
\begin{align} 
\mathcal{Z}  = \int D\phi D\pi \exp\left\{-\beta E[\phi,\pi]\right\},
\end{align}
and 
\begin{align}
		 E[\phi,\pi] & =\int d^3\vec{x} \left[ \frac{1}{2}
                   \pi(\vec{x})^2 + \frac{1}{2} (\nabla
                   \phi(\vec{x}))^2 + V(\phi)\right],
\end{align}
is the energy functional.

 \subsection{Computation of the rate}
 \label{sec:computationoftheratescalarfield}

 The computation of the escape rate for the scalar field is a
 generalization of the zero-dimensional flux-over-population method to
 stochastic field theory. The original extension of  the method to a
 $2N$-dimensional system has been performed by Langer  in
 Refs.~\cite{LANGER1967108,LANGER1969258}. 
 
\subsubsection{Setting up the problem} 

The flux-over-population method relies on similar ideas as in the
zero-dimensional case. The initial configuration  is a homogeneous and
static field located at the local minimum of the potential,
\begin{align}
	\phi_{i}(t_0)& = \phi^A_{i}, &&\pi_{i}(t_0) = 0, && \forall i.
\end{align}
We assume that, for large negative values of the field, the potential
is diverging and, on the other side,  there is a local maximum located
at $\phi^B$, as shown in {}Fig.~\ref{fig:ScalarPotentialEscape}. The
probability density  at time $t_0$ is a product of Dirac
delta-functions peaked at $\phi= \phi^A$ and $\pi=0$, 
   \begin{align}
 	P(\phi,\pi,t_0\mid\phi_0,\pi_0,t_0)&=\prod_{i=1}^{N^3}\delta
        [\pi_i]\cdot\delta[\phi_{i}-\phi^A].
 \end{align}
   After a sufficiently long time, the system is expected to be
   described by the equilibrium distribution given in
   Eq.~\eqref{eq:equilibriumdistributionfield}.  At its early stages,
   the evolution of the system implies an increasing probability to
   find the field on the other  side of the potential and, therefore,
   a flux of probability at the barrier. 
 
 The probability current is expected to go along the configuration
 with the minimal energy on the barrier ridge.  This defines the
 saddle-point configuration, which is found by taking the variation of
 the energy function
\begin{align}
	\delta E & = a^3 \sum_{i=1}^{N^3} \pi_i \delta \pi_i
        +\left[-\nabla^2 \phi_i+ V'(\phi_i) \right]\delta \phi_i.
\end{align}
We directly observe that the initial configuration is an extremum of
the energy. The next configuration that extremizes the  energy is
given by $\pi^S_i=0$ and $\phi^S_i$ that satisfies the saddle-point
equation,
\begin{align} 
	\label{eq:saddlepointequationdefinition}
	\nabla^2 \phi^S_i = V'(\phi^S_i),
\end{align}
and defines the saddle-point configuration. The exact form of the
solution $\phi^S$ is  {\it a priori} not obvious.  As stated in 
Sec.~\ref{sec:CHdefinitionoftheescapaeratefield}, a simple solution
is the homogeneous case where the  field is at the top of the
potential $\phi^B$, at each point of space. This trivial solution of
the saddle-point equation  is relevant in a situation where the volume
of space in consideration is finite. An example is the early Universe
where  this solution corresponds to the Hawking-Moss
instanton~\cite{HAWKING198235}, and the volume is a sphere of Hubble
radius.   In our case, the volume of space might be arbitrarily large.  It is fair to assume that the rate of escape, which has an exponential dependence on the volume of space that experiences hopping, will be strongly suppressed when considering a large or even infinite volume. We, therefore,  seek for a solution of the saddle-point equation where only a finite region of the space escapes, as already suggested in Ref.~\cite{zinn2002quantum}.
 
We might  try to find a
solution of Eq.~\eqref{eq:saddlepointequationdefinition} where the
field is homogeneously sitting at the initial  position $\phi^A$
everywhere except in some finite part, where it is climbing the
potential well. Using the rotational symmetry  and  writing the
saddle-point equation in spherical coordinates, we obtain
 \begin{align} 
	 \label{eq:saddlepointequationdefinitioninsphericalcoordinate}
	\frac{\partial^2 }{\partial r^2} \phi^S  +\frac{2 }{r}
        \frac{\partial }{\partial r} \phi^S = V'(\phi^S),
\end{align}
where, for simplicity, we are working in the continuum limit. The
boundary conditions are 
\begin{align}
	 \label{eq:saddlepointboundarycondition}
	\lim_{r\rightarrow\infty} \phi^S(r)& = \phi^A,&&
        \left. \frac{\partial }{\partial r} \phi^S \right|_{r=0}=0,
\end{align}
 where the second condition has been introduced to make sure the
 left-hand side of the saddle-point
 equation~\eqref{eq:saddlepointequationdefinitioninsphericalcoordinate}
 is finite at the center of the coordinates.  The equation can be
 interpreted as the equation of motion of a fictitious point particle,
 in an inverted potential  $-V$ and with a damping term. The
 overshoot/undershoot technique of Coleman~\cite{Coleman:1977py} shows
 that a  solution  only exists if the original minimum is a false
 vacuum. We should then consider separately the cases where  the
 initial minimum is a false (given by the dotted and dashed-dotted lines
 shown in {}Fig.~\ref{fig:ScalarPotentialEscape}) or a true
 (e.g., the dashed line shown in
 {}Fig.~\ref{fig:ScalarPotentialEscape}) vacuum.

In the case of a false vacuum at $\phi^A$, the saddle-point solution
satisfying
Eq.~\eqref{eq:saddlepointequationdefinitioninsphericalcoordinate}
exists and it is well understood. Let us consider the two limiting
cases. If the potential is unbounded from below, by continuity, there
must be  a field value $\phi^{0}>\phi^{U0}$, where the fictitious
particle starts at $r=0$ with zero velocity and reaches $\phi^A$ at
infinite radius.   Moreover, it has been shown in
Ref.~\cite{Linde:2005ht} that $\phi^0$ is of the order  of $\phi^{U0}$. In
the presence of a true vacuum at  $\phi^{TV}$,  the existence of a
solution is ensured by the overshoot/undershoot argument. If
$V(\phi(r=0)) > V(\phi^A)$, the fictitious particle does not  have
enough potential energy to climb the inverted potential up to
$\phi^A$, this is an undershoot. On the other hand, if $\phi(r=0)$ is
close  enough to $\phi^{TV}$, the fictitious particle can stay near
the true minimum until the damping term becomes negligible, since it
is suppressed  by $r$, and then it will overshoot. By continuity,
there is a field value to start at $r=0$ that satisfies $V(\phi(r=0))
< V(\phi^A)$ and  $\phi(r=0)<\phi^{TV}$  such that the fictitious
particle ends at $\phi^A$ at infinite radius. By these arguments, the
saddle-point configuration  is uniquely defined as well as the region of space that experiences hopping. Moreover, it has been
 shown by Coleman in Ref.~\cite{COLEMAN1988178} that the Hessian
matrix of the energy evaluated for this  configuration has only one
negative eigenvalue.  In the presence of an initial false vacuum, the escape rate  is therefore analogous to the rate of nucleation of a critical bubble of true vacuum, due to thermal fluctuations, but now with in addition the damping explicitly taken into account.

One of the main differences with nucleation is that the escape problem can be defined for an initial true vacuum at $\phi^A$. However, a proper definition of the escape rate in this case requires additional care. On the one hand, by comparing the initial and  the
equilibrium  distributions, it is fair to assume that there is a
probability flow at the potential barrier and, therefore, it should be
possible to define an  escape. On the other hand, the undershoot
argument forbids the existence of a solution of the saddle-point
equation. We will come back to this  issue at the end of this section, in Sec.~\ref{subsec:initialstableminimum},
and make some propositions for a well-defined escape problem. {}For
the moment, we simply assume that the  initial  position $\phi^A$ is a
false vacuum and proceed with the computation of the escape rate. Note that the fact that the escape rate depends on the initial minimum being global or local is a clear difference with the point particle case discussed in Sec.~\ref{sec:classicalpointparticlereview}. For a field, the shape of the potential beyond the barrier plays a role in the evaluation of the escape rate.

The flux-over-population method relies on the following assumptions: 
\begin{itemize}
	\item No sources or sinks lie in the neighborhood of
          the saddle-point configuration. This allows us  to write the
          FP  equation~\eqref{eq:FPEscalarfield} near the saddle point
          as  
\begin{widetext}
	\begin{align}
 		 & a^3\sum_{i=1}^{N^3} \left\{-\pi_i
          \frac{\partial}{a^3\partial
            \phi_i}+\frac{\partial}{a^3\partial \pi_i}\left[\eta \pi_i
            + a^3
            \sum_{k=1}^{N^3}\left[-\frac{\nabla^2_{ik}}{a^3}+
              \frac{V''(\phi^S_k)\delta_{ik}}{a^3}\right]
            (\phi_k-\phi_k^S)\right]+\frac{\Omega}{2}
          \frac{\partial^2}{ a^6\partial\pi_i^2}\right\} P(\phi,\pi)
          = 0,
	\end{align}
\end{widetext}
	using the expansion of the energy near the saddle point
	\begin{eqnarray}
		E[\phi,\pi]  &=&  E[\phi^S,\pi^S]  \nonumber \\ &+&
                \frac{1}{2} a^6 \sum_{i,j = 1}^{N^3}
                (\phi_i-\phi_i^S)\left[-\frac{\nabla^2_{ij}}{a^3}+
                  \frac{V''(\phi^S_i)\delta_{ij}}{a^3}\right](\phi_j-\phi_j^S)
                \notag\\ &+& \frac{1}{2}a^6\sum_{ij = 1}^{N^3}
                (\pi_i-\pi_i^S)\frac{\delta_{ij}}{a^3}(\pi_j-\pi_j^S)+\dots
                .
	\end{eqnarray}
	In the spirit of the vector-matrix notation defined above, we
        introduce the matrix $(e^S_{ij})$
		\begin{align}
		(e^S_{ij}) & = -
                  \frac{1}{a^3}\left(  \begin{array}{cc}
                    -{\nabla^2_{ij}}+{V''(\phi^S_k)\delta_{ij}} &  0
                    \\ 0 & \mathbb{1}  \end{array}\right) ,
	\end{align}
	which corresponds to the negative of the Hessian matrix of the
        energy evaluated at the saddle-point
        configuration.  In the context of field
          theory, the Hessian matrix is usually referred to as a fluctuation
          operator.

	\item Inside the well, near the minimum where the field is
          located initially, the system is thermalized,
	\begin{align}
		P(\phi \simeq \phi^A, \pi\simeq\pi^A) & \simeq
                P_0(\phi,\pi),
	\end{align}
	where $P_0$ is the equilibrium distribution.
	\item Beyond the saddle point, the probability density is
          strongly suppressed due to the presence of the sinks.
\end{itemize}

\subsubsection{The derivation of the probability density}
 
The computation of the flow of the probability current and the number
density relies  on the solution $P(\phi,\pi)$ of the FP equation with
the boundary conditions given above. This solution is derived using
the Kramers ansatz,
\begin{align}
	P(\phi,\pi) & = \zeta(\phi,\pi)P_0(\phi,\pi),
\end{align}
where $ \zeta(\phi,\pi)$ must be fixed to satisfy the boundary
conditions,
\begin{align}
	\zeta(\phi\simeq\phi^A,\pi\simeq\pi^A) & = 1 , && \zeta(\phi >
        \phi^S,\pi) \rightarrow 0.
\end{align}
The equation for $\zeta(\phi,\pi)$ is found by insertion in the FP
equation. In particular, near the saddle point, one finds
\begin{widetext}
\begin{align}
	\label{eq:equationforzeta}
	a^3\sum_{i=1}^{N^3} & \left\{-\pi_i
        \frac{\partial}{a^3\partial \phi_i}+\left[-\eta \pi_i + a^3
          \sum_{k=1}^{N^3}\left[-\frac{\nabla^2_{ik}}{a^3}+
            \frac{V''(\phi^S_k)\delta_{ik}}{a^3}\right]
          (\phi_k-\phi_k^S)\right]\frac{\partial}{a^3\partial
          \pi_i}+\frac{\Omega}{2} \frac{\partial^2}{
          a^6\partial\pi_i^2}\right\} \zeta(\phi,\pi)  = 0.
\end{align}
\end{widetext}
With the same arguments as in the zero-dimensional point particle case
and following the Kramers original proposal, it is assumed that
$\zeta(\phi,\pi)$ depends on a linear combination $u$ of the $\phi_i$
and $\pi_i$,
\begin{align}
	\zeta(\phi,\pi) & = \zeta(u),
\end{align}
with
\begin{equation}
u= a^3
\sum_{i=1}^{N^3}\left[U_i(\phi_i-\phi_i^S)+\bar{U}_i(\pi_i-\pi_i^S)\right],
\end{equation}
where $U_i$ and  $\bar{U}_i$  are the coefficients associated with
$\phi_i$ and $\pi_i$, respectively.  Similar arguments as the ones stated after Eq.\eqref{eq:Kramersassumption} for the zero-dimensional point particle case  justify this form of solution $\zeta(u)$. The following ansatz for $
\zeta(u)$,
\begin{align}
	\zeta(u) & = \frac{1}{\sqrt{2\pi k_B T}}
        \int_u^{\infty}dz\ \exp\left\{-\frac{ z^2}{2 k_BT}\right\},
\end{align}
satisfies the boundary conditions. To compute the coefficients $U_i$
and $\bar{U}_i$, we substitute $\zeta(u)$ in
Eq.~\eqref{eq:equationforzeta} and obtain
\begin{widetext}
	\begin{align}
		\label{eq:equationforUandUbardiscrete}
		a^3\sum_{i=1}^{N^3} &
                \left\{\left(U_i+\eta\bar{U}_i\right)\pi_i-\bar{U}_ia^3
                \sum_{k=1}^{N^3}\left[-\frac{\nabla^2_{ik}}{a^3}+
                  \frac{V''(\phi^S_k)\delta_{ik}}{a^3}\right](\phi_k-\phi_k^S)
                \right.\notag\\ & \left.+\ \eta \bar{U}_i^2  a^3
                \sum_{k=1}^{N^3} U_k(\phi_k-\phi_k^S) + \eta
                \bar{U}_i^2  a^3
                \sum_{k=1}^{N^3}\bar{U}_k(\pi_k-\pi_k^S) \right\}   =
                0.
	\end{align}
\end{widetext}
 At first sight, this equation seems
unpromising. {}Fortunately, it can be written in a simple form using
the vector-matrix notation that has been introduced
previously.  Defining the $(2N^3)$ vectors $(U\ \bar{U})^T$ and
$(\phi-\phi^S\ \pi-\pi^S)^T$ such that 
\begin{align}
	u & = (U\ \bar{U}) \cdot \left(  \begin{array}{c} \phi-\phi^S
          \\  \pi-\pi^S \end{array}\right)  \nonumber \\ &= a^3
        \sum_{i=1}^{N^3}
        \left[U_i(\phi_i-\phi_i^S)+\bar{U}_i(\pi_i-\pi_i^S)\right],
\end{align}
with the scalar product being defined as in
Eq.~\eqref{eq:scalarproductmatrixfield}, the equation for the
parameters $U_i$ and $\bar{U}_i$ becomes
\begin{align}
	\label{eq:matrixequationUUbar}
	& (U\ \bar{U}) \cdot M^T \cdot (e^S_{ij}) \cdot
        \left(  \begin{array}{c} \phi-\phi^S
          \\  \pi-\pi^S \end{array}\right)  \nonumber \\ & = \lambda
        (U\ \bar{U}) \cdot \left(  \begin{array}{c} \phi-\phi^S
          \\  \pi-\pi^S \end{array}\right),
\end{align}
where the scalar $\lambda$ is defined as
\begin{align}
	\lambda \equiv 	(U\ \bar{U}) \cdot M \cdot
        \left(  \begin{array}{c} U \\  \bar{U} \end{array}\right) & =
        a^3 \sum_{i=1}^{N^3}\eta\bar{U}_i\bar{U}_i.
\end{align}
The matrix equation~\eqref{eq:matrixequationUUbar} leads to the
eigenvalue equation for $ (U\ \bar{U})$,
\begin{align}
	 (U\ \bar{U}) \cdot M^T \cdot (e^S_{ij}) & = \lambda
  (U\ \bar{U}) , \label{eq:eigenvalueequationforu}
\end{align}
and the other term $ (U\ \bar{U})^T$ is a left eigenvector of the
matrix $M^T \cdot (e^S_{ij})$ with eigenvalue $\lambda$.  Combining
the definition of $\lambda$ and the eigenvalue equation, we find the
normalization condition 
\begin{align}
	1 =  (U\ \bar{U}) \cdot (e^S_{ij})^{-1}  \cdot
        \left(  \begin{array}{c} U
          \\  \bar{U} \end{array}\right). \label{eq:normalizationsf}
\end{align}
The  eigenvalue $\lambda$ is positive  by definition. The positivity
is, in fact, a direct consequence of the overall negativity  of the
exponent of $\zeta(u)$. This negative exponent has been chosen in
order to satisfy the boundary condition imposed by the  method, namely
the suppression of the probability distribution  beyond the
saddle point, and $\lambda$ is the only positive  eigenvalue of the
matrix $M^T \cdot (e^S_{ij})$.  Recall that $(e^S_{ij})$ is defined as
the negative of the Hessian of the  energy, evaluated exactly at the
saddle point.

\subsubsection{The probability density current and flux}  

Once we have obtained the probability density $P=\zeta P_0$ we are
ready to compute  the associated probability density current defined
in Eqs.~\eqref{eq:ProbabilityCurrentJ}
and~\eqref{eq:ProbabilityCurrentJbar}. After  some algebra, we find
\begin{align}
	J\zeta P_0 &  = \sqrt{\frac{k_B T}{2\pi }} M \cdot
        \left(  \begin{array}{c} U \\  \bar{U} \end{array}\right)
        \exp\left\{-\frac{ u^2}{2 k_BT}\right\} P_0,
\end{align}
and the probability flux $j$ is given by
\begin{align}
	j = & \  a^3 \sum_{i=1}^{2N^3}\int_{u=0} dS_i\ J_i(\phi,\pi)
        \notag\\ = & \ \frac{ \lambda}{2\pi \mathcal{Z}}
        \sqrt{\frac{k_B T}{2\pi }}  \exp\left\{-\beta
        E[\phi^S,\pi^S]\right\}  \nonumber \\ & \times \int D\phi D\pi
        \int {dk} \exp\left\{ik (U\ \bar{U}) \cdot
        \left(  \begin{array}{c} \phi-\phi^S
          \\  \pi-\pi^S \end{array}\right) \right\}\notag\\ &\times
        \exp\left\{\frac{\beta}{2} \left(  \begin{array}{c}
          \phi-\phi^S \\  \pi-\pi^S \end{array}\right)^T \cdot
        (e^S_{ij})  \cdot \left(  \begin{array}{c} \phi-\phi^S
          \\  \pi-\pi^S \end{array}\right)   \right\}.
\end{align}
We can diagonalize the matrix $(e^S_{ij}) $ by introducing the
rotation $S=(S_{ij})$ in field space to obtain
\begin{align}
	 & \left(  \begin{array}{c} \phi-\phi^S
    \\  \pi-\pi^S \end{array}\right)  = S \cdot \xi,  \\ & iku = ik
  (U\ \bar{U}) \cdot S^\dagger \cdot S  \cdot \left(  \begin{array}{c}
    \phi-\phi^S \\  \pi-\pi^S \end{array}\right) = ik \tilde{U} \cdot
  \xi,
\end{align}
where we have defined the vector $\tilde{U}$ as $S\cdot  \left(
U\  \bar{U}\right)^T$ and
\begin{align}
 	\left(  \begin{array}{c} \phi-\phi^S
          \\  \pi-\pi^S \end{array}\right)^T  \cdot (e^S_{ij})  \cdot
        \left(  \begin{array}{c} \phi-\phi^S
          \\  \pi-\pi^S \end{array}\right)  &  = a^3 \mu_1 \xi_1^2 -
        a^3 \sum_{l=2}^{2N^3}\mu_l\xi_l^2,
\end{align}
where all the scalars $\mu_l$ are defined as positive.\footnote{For the
  moment, we ignore the possibility of vanishing eigenvalues. We shall
  come back to them shortly.}  The only positive eigenvalue of
$(e^S_{ij})$ is $\mu_1$, all the other eigenvalues are
$-\mu_l$. Hence, we finally can write the flux $j$ as
\begin{align}
	j  =  & \frac{\lambda}{2\pi\mathcal{Z}} e^{-\beta
          E[\phi^S,\pi^S]} |\det (2\pi/ \beta)^{-1}
        E^{(S)}|^{-\frac{1}{2}},
\end{align}
where the matrix $E^{(S)} = -(e_{ij}^S)$ is the Hessian of the energy
at the saddle point, and that has only one negative eigenvalue. Since
this negative eigenvalue appears with a negative sign, it is the
magnitude of the determinant that enters the formula. The successive
integrations have been performed in the following order, first over
all the modes $l$ larger than 1, then over $k$ and finally over
$\xi_1$. 

\subsubsection{The zero modes} 

Due to the translation invariance of the saddle-point solution, there
are three eigenvalues in the associated  determinant that are exactly
zero and, therefore, must be treated separately upon the Gaussian
integration. {}For simplicity and in order  to agree with the
literature, we perform the analysis in the continuum space. {}First
of all, let us show that $\partial_{\vec{x}}\phi^S$,
$\partial_{\vec{y}}\phi^S$ and $\partial_{\vec{z}}\phi^S$ are
zero-modes. Considering  $\partial_{\vec{x}}\phi^S$ we have
\begin{eqnarray}
&&[-\nabla^2 + V''(\phi^S)]\partial_{\vec{x}}\phi^S   \nonumber \\ &&=
  -\partial_{\vec{x}}\nabla^2\phi^S +
  V''(\phi^S)\partial_{\vec{x}}\phi^S \nonumber \\ &&
  =-\partial_{\vec{x}}V'(\phi^S) + V''(\phi^S)\partial_{\vec{x}}\phi^S
  \notag\\ &&= -V''(\phi^S)\partial_{\vec{x}}\phi^S +
  V''(\phi^S)\partial_{\vec{x}}\phi^S =0.
\end{eqnarray}
To remove the zero-modes, we follow the procedure described in
~\cite{LANGER1969258} and ~\cite{Callan:1977pt}.
{}First of all, the determinant has its zero-eigenvalues removed and
becomes
\begin{align}
	 & |\det (2\pi / \beta)^{-1}[-\nabla^2 + V''(\phi^S_i)]|
  \nonumber \\ & \rightarrow |{\det}'(2\pi / \beta)^{-1}[-\nabla^2 +
    V''(\phi^S_i)]|,
\end{align}
with the prime denoting the removal of the vanishing
eigenvalues. Then, the integration over the zero-modes
$\partial_{\vec{x}}\phi^S$, $\partial_{\vec{y}}\phi^S$ and
$\partial_{\vec{z}}\phi^S$ becomes an integration over $d\vec{x}$,
$d\vec{y}$ and $d\vec{z}$, giving an overall volume factor
$\mathcal{V}$. {}Finally each change of variable from the zero-modes
to  $\partial_{\vec{x}}\phi^S$, $\partial_{\vec{y}}\phi^S$ and
$\partial_{\vec{z}}\phi^S$ to $d\vec{x}$, $d\vec{y}$ and $d\vec{z}$
leads to a Jacobian. {}For example, for the mode
$\partial_{\vec{x}}\phi^S$ we have
\begin{align}
	 \left[\int d^3\vec{x}\ \left(\frac{\partial{\phi}^S}{\partial
             x}\right)^2\right]^{1/2}.
\end{align}
The Jacobian is identical for each zero-mode since
\begin{align}
\!\!\!	 \int d^3\vec{x}\ \left(\frac{\partial{\phi}^S}{\partial
  x}\right)^2 &=\int
d^3\vec{x}\ \left(\frac{\partial{\phi}^S}{\partial y}\right)^2  = \int
d^3\vec{x}\ \left(\frac{\partial{\phi}^S}{\partial z}\right)^2,
\end{align}
where we used the rotation-symmetry of the saddle-point solution. We
then have
\begin{align}
	\int d^3\vec{x}\ \left(\frac{\partial{\phi}^S}{\partial
          x}\right)^2 &= \frac{1}{3}\int
        d^3\vec{x}\ \left(\nabla{{\phi}^S}\right)^2.
\end{align}
Hence, there is an overall factor multiplying the rate coming from the
Jacobian and given by
\begin{align}
	\left[ \frac{1}{3}\int
          d^3\vec{x}\ \left(\nabla{{\phi}^S}\right)^2\right]^{3/2}.
\end{align}
A quick dimensional check tells us that removing the three eigenvalues
from the determinant increases  the dimension by $3/2$. The overall
volume factor has a dimension of $-3$ and the Jacobian $3/2$, exactly
compensating the removal  of the zero-eigenvalues.

\subsubsection{Population inside the well} 

The last missing piece is to account for the population inside the
well. This is obtained using the condition  that the system be
thermalized near the minimum of the potential and by expanding the
energy function around the configuration $(\phi^A,\pi^A)$,
\begin{eqnarray}
\lefteqn{	E[\phi,\pi] = E[\phi^A,\pi^A] } \\ \nonumber \\ &&+
\frac{1}{2} a^6 \sum_{i,j = 1}^{N^3}
(\phi_i-\phi_i^A)\left[-\frac{\nabla^2_{ij}}{a^3}+\frac{V''(\phi^S_i)
    \delta_{ij}}{a^3}\right](\phi_j-\phi_j^A)
\notag\\ &&+ \frac{1}{2}a^6\sum_{ij = 1}^{N^3}
(\pi_i-\pi_i^A)\frac{\delta_{ij}}{a^3}(\pi_j-\pi_j^A)+\dots .
\end{eqnarray}
The population $n_A$ inside the well is found to be given by
\begin{align}
	n_A & = \int D\phi D\pi P_0   \nonumber \\ &=
        \frac{1}{\mathcal{Z}}  e^{-\beta E[\phi^A,\pi^A]} [\det (2\pi
          / \beta)^{-1} E^{(A)}]^{-\frac{1}{2}},
\end{align}
where the matrix $E^{(A)}$ is the Hessian of the energy of the initial
configuration at $\phi^A$ and all eigenvalues are positive.

\subsubsection{The escape rate expression} 

The ratio of the flux $j$ over the number density $n_A$, taking into
account the zero-modes, gives the escape  rate $k$ for a scalar field
per unit volume,
\begin{align}
	\frac{k}{\mathcal{V}}  &=   \frac{\lambda}{2\pi} \left[
          \frac{1}{3}a^3\sum_{i=1}^{N^3}\
          \left(\nabla{\phi_i^S}\right)^2\right]^{3/2}
	\nonumber \\ & \times \left[\frac{ \det[ (2\pi / \beta)^{-1}
              E^{(A)}]}{|\det' [(2\pi / \beta)^{-1}
              E^{(S)}]|}\right]^{1/2} e^{-\beta
          \left[E(\phi^S,\pi^S)-E(\phi^A,\pi^A)\right]} .
	\label{krate}
\end{align}
Let us consider the different contributions to the rate. In the
exponent in Eq.~\eqref{krate} we have
\begin{align}
	& E(\phi^S,\pi^S)-E(\phi^A,\pi^A) \nonumber \\ & =a^3
  \sum_{i=1}^{N^3}  \frac{1}{2} (\nabla \phi^S_i)^2 +
  V(\phi^S_i)-V(\phi^A_i),
\end{align}
which corresponds to the activation energy, i.e.,   the difference
between the energy of the saddle-point configuration with respect to the
initial configuration. Since the initial configuration is homogeneous
and only a difference of potential energy enters the rate formula,  we can
safely shift the potential to have $V(\phi^A_i)=0$. The determinants
in Eq.~\eqref{krate} can be written as
\begin{align}
	\det [(2\pi / \beta)^{-1}E^{(A)}] & = \det[ (2\pi /
          \beta)^{-1}(-\nabla^2 + V''_A)],
\end{align}
where $V''_A$ is the second derivative of the potential at the initial
minimum and
\begin{align}
	|{\det}' [(2\pi / \beta)^{-1}E^{(S)}]| & = |{\det}' (2\pi /
        \beta)^{-1}[-\nabla^2 + V''(\phi^S)]|,
\end{align}
where the field configuration entering the operator is the
saddle-point solution. 

Substituting the above equations into Eq.~\eqref{krate}, we then find
that the escape rate per unit volume in the continuum limit is
\begin{align}
	\label{eq:finalresultescaperate}
	\frac{k}{\mathcal{V}}  & =  \frac{\lambda}{2\pi}\left[ \frac{
            \beta}{6 \pi }\int
          d^3\vec{x}\ \left(\nabla{{\phi}^S}\right)^2\right]^{\frac{3}{2}}
        \nonumber \\ & \times \left[\frac{\det [-\nabla^2 + V''_A]}{
            |{\det}' [-\nabla^2 + V''(\phi^S)]|}\right]^{\frac{1}{2}}
        e^{-\beta \int d^3\vec{x} \left[\frac{1}{2} (\nabla \phi^S)^2
            + V(\phi^S)\right]},
\end{align}
which, in conjunction with our derivation, 
is the main result of this paper. The constant $\lambda$
appearing in the above result is sometimes referred to as the dynamical
prefactor and the ratio  of determinants as the statistical
prefactor~\cite{LANGER1969258,RevModPhys.62.251,PhysRevA.8.3230,PhysRevD.46.1379}. The
explicit expressions of these factors depend on the saddle-point
configuration $\phi^S$. It should be noted that Eq.~\eqref{eq:finalresultescaperate}  
has many similarities to the corresponding expression for the particle case Eq.~\eqref{eq:CH4fopratepointparticlefinal}.
This should not be too surprising since escape rates are usually in the form of a statistical and a dynamical prefactor multiplying an exponential of the free energy associated with the escape. The difficulty lies in specifying the different prefactors and the free energy of the escape configuration. In particular, the dynamical prefactor $\lambda$ in Eq.~\eqref{eq:finalresultescaperate} does not take such a simple form as in Eq.~\eqref{eq:lambdappc}. The ratio of determinants and the Jacobian associated with the zero modes, which both form the statistical prefactor, are common in field theories.  Finally, it of course must be kept in mind that the expression in the exponential arises from an integral over the whole field configuration.  Methods specific to such a system and not just a point particle need to be utilized when evaluating, as is evidenced by the saddle-point configuration used above, which would not be possible for the point particle case.

We choose to present here the most  general form of the escape rate. An explicit estimation of the rate, in particular the prefactors and the exponent, is possible once a potential has been specified.   A discussion about the methods to estimate the rate in a practical case, for example the thin-wall approximation,  is given in Sec.~\ref{sec:explicitcomputation}.

\subsubsection{The initial stable minimum} 
\label{subsec:initialstableminimum}
The last case left to consider is when $\phi^A$ in
{}Fig.~\ref{fig:ScalarPotentialEscape} is a true vacuum. As described
above,  the saddle-point
equation~\eqref{eq:saddlepointequationdefinitioninsphericalcoordinate}
does not have any solution. However,  in the presence of fluctuation
and dissipation dynamics, it is fair to assume, in any given
realization  of the noise, that the field starts to climb the
potential and probes the other side of the well, even if it will
likely come  back to the original side. Moreover, as noted already, the comparison between the initial
probability distribution, which is a Dirac delta function
peaked at $\phi^A$ at each point in space, and the
equilibrium distribution, that probes both sides of the well,
implies a flow of probability through the maximum of
the potential. These two
arguments suggest that the escape  problem for an initial true vacuum
might still be defined. The rate will simply indicate how likely it is
to have a  region of space that passes the barrier.  Let us formulate
some propositions for a meaningful definition for this case. 

The first possibility for treating the present case is to consider a
finite volume  $\mathcal{V}$ of space and use the saddle-point
solution $\phi^S = \phi^B$ at each point in the volume. The activation
energy will be given by $E = \mathcal{V} \Delta V$. This is the
simplest generalization of the zero-dimensional case but  it is
dependent on the volume in consideration. Moreover, it can lead to an
underestimate of the rate since, instead of  waiting at the top of the
potential, the field can fall on the other side and attract the
neighboring points without  any additional energy. 

The method of reactive flux, described, for example, in the
review~\cite{RevModPhys.62.251}, might be helpful in the  derivation
of the escape rate for an initially true minimum. At equilibrium, the
ratio of particle densities in the  wells is equal to the ratio of
the rates between the two minima. Since the equilibrium distribution
and the rate from  a false to a true vacuum are known, the transition
rate from an initial true vacuum can be extracted. It is reasonable
to assume that, at equilibrium, the activation rate derived with the
method of reactive flux will be smaller than the  true escape
rate. However, this method also allows to study further the approach
to equilibrium by defining a relaxation  rate, from an initial
out-of-equilibrium distribution.

Alternatively, we can consider an approximated case, where, in
{}Fig.~\ref{fig:ScalarPotentialEscape}, the false minimum on the
right-hand side  is replaced by  a true minimum, due to a modification
of the potential beyond the maximum. {}For example, a minimal
situation could be a  new true minimum, almost degenerate with
$V(\phi^A)$. A saddle-point configuration is  well defined and the
rate is given  by Eq.~\eqref{eq:finalresultescaperate}. Moreover, the
saddle-point configuration will naturally define the typical size of
the region of space that experiences hopping. As in the previous case,
the escape rate might be underestimated. However,  it is fair to
expect that the main contribution to the escape time is given by the
climbing of the potential well, which  corresponds to the part of the
potential that is not modified.

The last possibility is the construction of a saddle-point configuration
using an analytic continuation. It was not possible  to obtain a
solution of
Eq.~\eqref{eq:saddlepointequationdefinitioninsphericalcoordinate},
where the field is at  $\phi^{FV}$ at $r=0$ and respecting the
boundary condition~\eqref{eq:saddlepointboundarycondition}. One can
imagine giving  an initial imaginary velocity to the field, which
would then allow for the climb. This kind of solution has been
studied  in the context of
tunneling~\cite{Bonini:1999cn,Ankerhold:2007,Bender:2010nu,Anderson:2011fn,Harada_2017}. However,
this goes  beyond the scope of this work, and we leave it for a future
analysis.

 \section{Discussion of the result}
\label{discussion}

Let us present here a comparison between the result we have obtained
for the escape rate, given by Eq.~\eqref{eq:finalresultescaperate}, and the
related problem of quantum tunneling  at a sufficiently high
temperature, where thermal effects dominate. The similarities between
the two results provide some insights  about the methods needed for an
explicit evaluation of the escape rate, once a potential has been
specified.

 \subsection{Comparison with quantum tunneling at finite temperature}
 \label{subsec:comparisonLanger}
 
 Quantum tunneling of a scalar field is a well-studied problem and
 plays a significant role in the study of first-order phase
 transitions and in the stability of false vacua. The problem has been
 solved for quantum field theory by Callan and  Coleman at
 zero temperature~\cite{Coleman:1977py,Callan:1977pt} and later
 extended to finite temperatures by   Linde~\cite{LINDE1983421}. The
 result of Ref.~\cite{LINDE1983421} is particularly interesting for
 the current analysis since, for sufficiently  high temperatures where the
 thermal fluctuations dominate over the quantum fluctuations, it recovers the result of Langer for classical nucleation~\cite{LANGER1967108,LANGER1969258}. In this
 regime, it is fair to expect  some similarities between the tunneling
 and the escape rates. 
 
The quantum tunneling rate per unit volume, at finite temperature and
when thermal fluctuations are dominant, is given by
\begin{align}
	\label{eq:LindeFTQT}
	\frac{\Gamma (T)}{\mathcal{V}}   =&\  T
        \left(\frac{\mathcal{S}_3(\phi^S,T)}{2\pi
          T}\right)^{\frac{3}{2}} \left[\frac{\det[-\nabla^2 +
              V''_A]}{ |{\det}' [-\nabla^2 +
              V''(\phi^S)]|}\right]^{\frac{1}{2}}  \nonumber \\ &
        \times \exp\left\{-\mathcal{S}_3(\phi^S,T)/T\right\},
\end{align}
 where the action $\mathcal{S}_3$ is defined as 
 \begin{align}
 	\mathcal{S}_3(\phi,T) & \equiv  \int d^3\vec{x}\left[
          \frac{1}{2} (\nabla \phi)^2 + V(\phi,T)\right],
 \end{align}
 and $\phi^S$ is a solution of 
  \begin{align} 
  	\label{eq:seFTQT}
	\frac{\partial^2 }{\partial r^2} \phi^S  +\frac{2 }{r}
        \frac{\partial }{\partial r} \phi^S = V'(\phi^S,T),
\end{align}
 where $V(\phi,T)$ is the temperature-dependent effective potential.
 
 Comparing with the escape problem, and assuming identical
 potentials,\footnote{To be more precise, we assume that the potential
   of the escape rate $V(\phi)$ is equal to the effective potential
   $V(\phi,T)$ at a fixed value of $T$.} we immediately notice that
 the field configurations entering the two rates are the same,
 Eq.~\eqref{eq:saddlepointequationdefinitioninsphericalcoordinate} for
 the case of the escape problem  and Eq.~\eqref{eq:seFTQT} given
 above. This similarity implies that the ratio of determinants and the
 exponential term are identical in the escape rate
 Eq.~\eqref{eq:finalresultescaperate} and in the tunneling rate
 Eq.~\eqref{eq:LindeFTQT}, respectively. Using the  argument of
 Coleman~\cite{Coleman:1977py,coleman_1985}, that the action
 $\mathcal{S}_3$ is invariant under an infinitesimal scale
 transformation of the solution $\phi^S$, we obtain
  \begin{align}
 	\mathcal{S}_3(\phi,T) & = \frac{1}{3}  \int
        d^3\vec{x}\  (\nabla \phi)^2 ,
 \end{align}
 which is precisely the term given by the Jacobian in the escape rate
 problem.
 
 The crucial difference between the escape and the quantum tunneling
 rates lies in the prefactors. In particular,  the escape rate
 predicts a factor of $\lambda/ 2\pi$ replacing the temperature. We
 interpret this difference as follows.  {}First of all, the escape
 problem, even if closely related, is not defined exactly  as the
 transition rate due to tunneling  effects. A comparable, but not
 identical, rate should emerge. Moreover, to derive the escape rate,
 we used the framework  of stochastic field theory, where the strength
 of the noise and the damping appear explicitly. One naturally expects
 the  damping to play a role in the final result, in particular within
 the dynamical prefactor $\lambda$ in Eq.~\eqref{eq:finalresultescaperate}.  On the contrary, the temperature that appears in the prefactor of~\eqref{eq:LindeFTQT} is an approximation which relies on dimensional grounds. In the approach of Ref.~\cite{LINDE1983421}, the properties characterizing the medium, for example the viscosities, are not taken into account. When the properties of the medium are considered~\cite{PhysRevD.46.1379,Venugopalan:1993vk,Csernai:2002bu}, a dynamical prefactor is expected in the rate. 
 
 It is however remarkable that
 the two rates computed with different methods, the stochastic field
 theory for the escape problem and the path integral formalism of
 quantum field theory for  tunneling, have so much in common. The
 escape rate only takes  into account the thermal fluctuations and is
 valid for  arbitrarily small temperatures. It is a strong support for
 Eq.~\eqref{eq:finalresultescaperate}  that the tunneling rate, in the limit where the  thermal
 fluctuations dominate, mostly recovers the escape rate.

 \subsection{Towards an explicit evaluation of the escape rate}
 \label{sec:explicitcomputation}
 
 In general, once  a potential has been specified, a complete
 derivation of the escape rate, Eq.~\eqref{eq:finalresultescaperate},
 requires numerical methods,  as for example in Ref.~\cite{Moore:2000jw}. However, exploiting the similarities with
 the quantum tunneling rate, we can use the techniques developed for
 the latter to  provide some guidance on the explicit derivation of
 the escape rate. This holds even if the derivations of both rates, as stated before, are based on completely different methods, path integrals for tunneling and a stochastic approach for the escape rate. Let us consider the exponent, the dynamical and
 statistical prefactor terms appearing in our final result
 Eq.~\eqref{eq:finalresultescaperate}, separately. Recall that, in
 general, it is sufficient to know  the order of magnitude of the
 prefactors, the rate being mainly dictated by the exponential.

\subsubsection{The exponent term} 

The evaluation of the exponent term in
Eq.~\eqref{eq:finalresultescaperate} requires the solution of the
saddle-point
equation~\eqref{eq:saddlepointequationdefinitioninsphericalcoordinate},
which, in general, is obtained numerically. However, two cases have
been identified where an analytical treatment is
possible~\cite{LINDE1983421,Linde:2005ht}. In the thin-wall
approximation, the potential has two  minima that are almost
degenerate. The saddle-point configuration has the form of a bubble
of true vacuum. Going along the radial direction, $\phi^S(r)$ is
initially almost constant and close to $\phi^{TV}$. This corresponds
to the interior of the bubble.  The field solution then bounces to
$\phi^A$, which defines the wall of the bubble. The critical radius of the
bubble is found by minimizing the energy.  It has been shown in~\cite{LINDE1983421,Linde:2005ht} that the
exponent becomes 
 \begin{align}
	 \int d^3\vec{x}\left[ \frac{1}{2} (\nabla \phi^S)^2 +
           V(\phi^S)\right] & =
         \frac{16\pi}{3\epsilon^2}\left(\int_{\phi^{TV}}^{\phi^A}
         d\phi\sqrt{2V(\phi)}\right)^3,
 \end{align}
 where $\epsilon$ is the difference between the false and true vacua,
 and the integral on the right-hand side is evaluated in the limit
 where $\epsilon$ vanishes.  The other situation where an analytical
 treatment is possible is when the potential difference between the false
 and true vacua is much larger than the barrier height. The potential
 can be approximated by a cubic or a quartic polynomial function in
 the field, leading to exact solutions. 
 
\subsubsection{The statistical prefactor} 

The exact evaluation of ratios of determinants in field theory is, in
general, an involved task. Recent discussions on some  analytical
approaches to this problem can be found in ~\cite{Gleiser:1993hf,
  Dunne:2007rt,Andreassen:2017rzq}. {}For the evaluation of the escape
rate just as for the tunneling case, as stated in ~\cite{LINDE1983421,Linde:2005ht}, it is
sufficient to have only a rough estimate of this prefactor. 
Dimensional analysis shows that the square root of the ratio of
determinants has dimension $m^3$ corresponding to the removal of the
three eigenvalues in the denominator. Therefore, we can write
 \begin{align}
	 \left[\frac{\det[-\nabla^2 + V''_A]}{ |{\det}' [-\nabla^2 +
               V''(\phi^S)]|}\right]^{\frac{1}{2}} & \sim
         \mathcal{O}\left(\phi^3, (V'')^{3/2},r^{-3},T^3\right),
 \end{align}
where the quantities on the right-hand side (apart from the temperature)
should be understood as mean values. In general, $\phi^3$, $
(V'')^{3/2}$, and $r^{-3}$ are of the same order of magnitude and
should be compared with the temperature to find the dominant
contribution. This is different from the case of quantum tunneling at a
finite temperature, where the temperature is expected to dominate in
the statistical prefactor.

\subsubsection{The dynamical prefactor} 

The dynamical prefactor $\lambda$ has been defined in
Eq.~\eqref{eq:eigenvalueequationforu} as the unique positive
eigenvalue of the matrix $M^T \cdot (e^S_{ij})$. The eigenvalue
equation for $\lambda$ can be written as
 \begin{align}
 		\left[\frac{\partial^2 }{\partial r^2}   +\frac{2 }{r}
                  \frac{\partial }{\partial r}  - V''(\phi^S)\right]
                v(r) & = \lambda(\lambda+\eta) v(r).
 \end{align}
 We observe that $\lambda$ has a dependence on the dissipation
 coefficient $\eta$. As usual, an analytical solution of the
 eigenvalue equation is not  possible, in particular, since it
 requires the knowledge of the saddle-point configuration
 $\phi^S(r)$. There exists, however, certain situations where an
 approximate result might be obtained, for example in the thin-wall
 approximation discussed above. Useful discussions on this problem can
 be found in Refs.~\cite{PhysRevA.8.3230,PhysRevD.46.1379,Venugopalan:1993vk,Csernai:2002bu}. 

 \section{Possible Applications}
\label{sec:applicationsforcosmologyandbeyond}
 
Here we will identify some suggested applications for the derived result
for the escape problem in field theory.  In fact, there can be applications
in any situation involving a phase transition.  In high-energy
physics this could be in the context of cosmology as well as in
heavy-ion collision experiments, and at low energy in
condensed matter systems. In all these cases one
can find applications where the escape problem defined here
plays a relevant role. A particular interest  is to look at scenarios
where the escape rate provides an alternative mechanism to quantum
tunneling.  We also identify problems where the methods, developed here in order to derive the escape rate, provide
an alternative approach. Since the aim of the current analysis is a formal
definition and a solution of the Kramers problem, we restrict to a
general description of these applications. A deeper analysis is left
for future works.

\paragraph{Phase transitions and topological defects:}
A concrete situation where the escape rate  becomes significant is in
the study of out-of-equilibrium systems, in particular, during a
first-order phase transition. Our analysis is well suited to
investigate the approach to equilibrium. We can imagine, for example,
the situation of an initially quadratic effective potential that is
developing another local minimum. The second minimum is, at first, a
false vacuum before becoming the true vacuum of the potential. The
escape rate provides the necessary tools to study the evolution of the
FP probability distribution between the old and the new equilibrium
distributions. 
 
 Phase transitions are often associated with the formation of
 topological defects~\cite{Kibble:1976sj,Vilenkin:2000jqa}. {}Fluctuation and dissipation
 dynamics can influence their creation, in particular in a
 second-order phase transition, where the height of the potential
 barrier is suppressed at the beginning of the transition. These
 effects might also play an important role in crossover
 transitions. In the special case of embedded defects~\cite{Vachaspati:1992pi, Zhang:1997is,Vachaspati:1992fi,Vachaspati:1992jk}, the possibility
 for the field to escape would have some consequences on the stability
 of the configuration. Examples of realistic stable embedded defects are known~\cite{Berera:2016vhw}. The escape rate should, therefore, be related to the destruction probability of  such a stable embedded configuration.

 \paragraph{Landscape of metastable minima:}
One of the most interesting features of the escape problem is the
hopping of the field over the potential barrier. Naively, considering
a potential with two minima that are almost degenerate, the escape
rate between the false and the true vacua should not be sensibly
different from the rate between the true and the false vacua. {}For
these reasons, the escape rate could be relevant in theories that
contain several nondegenerate minima, in particular, in order to
compute the probability for a finite part of space to evolve from one
minimum to the next.   One can imagine,
for example, a situation with two possible directions to
diffuse. In one of them, there is a large potential barrier but
a minimum at a lower energy beyond the well.
In the other direction, the potential barrier is smaller
but the next minimum is at a higher energy.
Quantum tunneling could only be applied to the first
case but the escape mechanism is applicable in both cases.

Such a situation arises in
string theories, which contain many
metastable vacua~\cite{Douglas:2003um}. This framework is called the
string landscape~\cite{Susskind:2003kw}. The question of how a vacuum
is selected is of particular interest. Our mechanism precisely allows
for the hopping from one vacuum to the next one. Moreover, the
Hagedorn temperature~\cite{Atick:1988si,Bowick:1989us}, usually
associated with string theories, could be the origin of the
fluctuation and dissipation dynamics. Such an analysis might require a
generalization of our work to take into account gravitational effects.

An active field of research in condensed matter physics concerns the
glass transition~\cite{Bertier2011}, corresponding to a phase
transition between a liquid and a glassy state. The phenomenology of
glassy systems can be described by an $N$-body system in a potential with several
metastable minima, called the potential energy
landscape~\cite{Stillinger1935,Debenedetti:2001aa}. The escape rate
provides a mechanism to probe the different minima. A generalization
of our analysis
to a nonrelativistic field would be needed in this case. 
 
\paragraph{Stochastic inflation:}
The stochastic formulation of inflation was introduced by
Starobinsky~\cite{STAROBINSKY1982175,Starobinsky:1986fx} as a
framework to study the dynamics of a quantum scalar field during
inflation.  The field is split into two parts, the long-wavelength part (coarse grained) and short-wavelength quantum fluctuations. The backreaction of the
quantum fluctuations on the coarse grained part  is
parametrized as a stochastic noise. The equation of motion of the
inflaton becomes a Langevin equation. The framework is particularly
relevant in the computation of correlation functions of the inflaton
field~\cite{Vennin:2015hra}.
 
 In general, the noise is assumed to be homogeneous and the problem
 reduces to the zero-dimensional case described in
 Sec.~\ref{sec:classicalpointparticlereview}. This approach  considers
 only the fluctuations that can lift an entire Hubble sphere. If, on
 the other hand, we imagine that the backreaction coming from the
 quantum fluctuations is inhomogeneous, the formalism developed for
 the escape rate is particularly useful. One can also think about
 different regions of space that evolve along different directions in
 the inflationary potential.

 \paragraph{Stochastic quantization:}
 The stochastic approach of quantum mechanics was first proposed by Nelson
 in ~\cite{PhysRev.150.1079} and then extended to fields by Parisi
 and Wu in ~\cite{Parisi:1980ys}. The main idea relies on the fact
 that the generating functional of Euclidean field theories is related
 to the equilibrium limit of a statistical system coupled to a heat
 reservoir. The temperature of the heat bath is chosen to match the
 Planck constant. The evolution of the system plus reservoir is in a
 fictitious time and the equilibrium is reached when this extra  time
 direction goes to infinity. This method for modeling quantum field
 theory is particularly useful for numerical simulations, such as in
 lattice field theory~\cite{DAMGAARD1987227}.
  
  The stochastic field theory introduced  for the derivation of the
  escape rate is formally equivalent to the formalism describing
  stochastic quantization. The only  difference is the dimension of
  space. The formalism described in Sec.~\ref{sec:secstocasticFT} can
  be seen as a three-dimensional Euclidean field theory coupled to a
  heat bath, whereas the stochastic quantization considers a
  four-dimensional Euclidean field theory and an extra time
  dimension. In the language of stochastic quantization, in particular,
  using the identification $\hbar = k_B T$, we can directly write the
  escape rate as
\begin{align}
\label{eq:quantumescaperate}
	\frac{k}{\mathcal{V}} & = \frac{\lambda}{2\pi}\left[ \frac{
            \mathcal{S}_4}{2 \pi \hbar}\right]^{2}\left[\frac{\det
            [-\Box + V''_A]}{ |{\det}' [-\Box +
              V''(\phi^S)]|}\right]^{\frac{1}{2}}
        e^{-\mathcal{S}_4(\phi^S)/\hbar},
\end{align}
where 
  \begin{align}
	 \mathcal{S}_4(\phi) & \equiv  \int d^4\vec{x}\left[
           \frac{1}{2} (\nabla \phi)^2 + V(\phi) \right],
\end{align}
and $\phi^S$ is the saddle-point configuration.  If the system is initially in a false vacuum, the quantum escape rate~\eqref{eq:quantumescaperate} defined for quantum fluctuations gives a quantum nucleation rate, which should be equivalent to quantum tunneling in the usual quantization.  A similar treatment
as in Sec.~\ref{subsec:comparisonLanger}  should be performed to
compare this result with the quantum tunneling rate at
zero-temperature, computed in
Refs.~\cite{Coleman:1977py,Callan:1977pt},  and to study how the two results agree. We leave the analysis of the quantum escape rate open for future works.

  \section{Conclusion}
\label{conclusions}

  In this work, we have proposed a definition and a solution of the
  Kramers problem for a scalar field theory. Using the framework of
  stochastic field theory, we have studied the probability for a
  scalar field to escape a potential well due to thermal
  fluctuations. The field theory character of the problem complicates
  the definition of the escape configuration.  Unlike the
  zero-dimensional point particle case, we have learned that the shape
  of the potential, beyond the local maximum, influences the  rate.
  Two situations have been identified that need to be treated separately,
  when the initial minimum corresponds to a true or a false
  vacuum. Using a generalization of the  flux-over-population method
  to a field, we have derived a full solution of the escape problem
  from a metastable vacuum and stated some directions to address the
  case of an initial true vacuum.
  
The main result of our analysis is the expression of the escape rate,
Eq.~\eqref{eq:finalresultescaperate}. A comparison with the quantum
tunneling rate, in the limit where the thermal fluctuations dominate,
shows that the two rates have much in common. These similarities
provide strong support for our result, in particular, since both
rates are computed from different approaches. The rates are, however,
not identical. This is not surprising, since the two problems, even if
related, are not exactly the same. In particular, the escape rate
explicitly takes damping effects into account. Nevertheless, the
well-studied framework of quantum tunneling provides some useful
techniques for an explicit evaluation of the escape rate, once a
potential is fixed. In this work, we have taken specific advantage of
that. It is remarkable that the derivation presented in this paper
also encompasses the Hawking-Moss instanton. This solution naturally
emerges from the flux-over-population method and can be studied
within the framework presented here.

Beyond the formal interest of the Kramers problem in field theory, we
have identified several concrete situations, in cosmology, particle
physics and condensed matter physics, where the escape rate is
relevant. Out-of-equilibrium scenarios, for example during a
transition between two nondegenerate vacua are natural candidates. In
cosmology, phase transitions and the formation of topological defects,
as well as stochastic inflation are various applications. The string
landscape and the glass transition present a favorable environment for
an escape mechanism.  On a more formal level, the analogy with the
stochastic quantization might shed  new light on both the
interpretation of the escape problem and on the meaning of the
stochastic approach of quantum mechanics.  A deeper analysis of these
directions will require further work.

\acknowledgments  A.B. is supported by STFC.  J.M. is supported by
Principal's Career Development Scholarship and Edinburgh  Global
Research Scholarship.  B.W.M. and
R.O.R. thank the Brazilian funding agencies CAPES, CNPq and  FAPERJ
for financial support. B. W. M. and R. O. R. are partially
supported by research grants from Conselho Nacional de
Desenvolvimento Cient\'{\i}fico e Tecnol\'ogico (CNPq), under
Grants No. 431.796/2016-5 (B. W. M.) and No. 302545/2017-4
(R. O. R.) and also by Funda\c{c}\~ao Carlos Chagas Filho
de Amparo \`a Pesquisa do Estado do Rio de Janeiro
(FAPERJ), under Grants No. E-26/202.649/2018 (B. W. M.) and
No. E-26/202.892/2017 (R. O. R.).

\appendix

\section{Mean first passage time}
\label{ap:MFPT}

\subsection{Definition of the MFPT over the barrier}

An alternative derivation of the escape rate is achieved with the
method of the mean first passage time. The first passage time (FPT)
is defined as the time the particle takes to leave a domain
$\mathcal{D}$ for the first time. In our case, it corresponds to the
time needed for the  particle initially at $x_A$, to pass over the
maximum at $x_B$ as depicted in Fig.~\ref{fig:1DPotentialEscape}. Since the forces acting on the particle are random
and the dynamics  not deterministic, the FPT is different for each
realization. One can, however, define the MFPT as the average of the
FPT and estimate  the escape rate as its inverse.  

A formal definition of the problem relies on the introduction of the
survival probability $S(t \ | \ x_0,v_0,t_0)$. It corresponds to  the
probability that the particle is still in $\mathcal{D}$ after a time
$(t-t_0)$, while being initially at position $x_0$ with velocity
$v_0$. In our case, the domain is the $A$-well where
$x\in(-\infty,x_{Si}]$, where the upper limit of the domain, $x_{Si}$,
        is a point  chosen to be near, but beyond, the maximum, to
        ensure the passing of the particle. The survival probability
        is defined as
\begin{align}
	S(t\mid x_0,v_0,t_0)&=\int_\mathcal{D} dxdv\ P(x,v,t\mid
        x_0,v_0,t_0)  \nonumber \\ &=
        \text{Prob}\left[T(x_0,v_0)>(t-t_0) \right]  \notag\\ &=
        \int_{(t-t_0)}^\infty dt\ f(t\mid x_0,v_0),\label{eq:SurCL}
\end{align}
 where $T(x_0,v_0)$ is the FPT starting at $x_0$ with initial velocity
 $v_0$ and $f(t\mid x_0,v_0)$ is the probability distribution for
 $T(x_0,v_0)$. The above relation is motivated by the following
 reasoning. The probability to be in the domain at time $t$ is the
 same as the probability of having a first passage time larger than
 $(t-t_0)$. 
 
{}From Eq. \eqref{eq:SurCL}, we deduce the following relation
between $S(t \mid x_0,v_0,t_0)$ and $f(t\mid x_0,v_0)$,
 \begin{align}
 	f(t\mid x_0,v_0)&=-\frac{\partial S(t\mid
          x_0,v_0,t_0)}{\partial t}.
 \end{align}
 The moments  $\langle T^n \rangle$ of the FPT are defined as
 \begin{align}
 	\langle T^n \rangle &  \equiv \int_{t_0}^\infty dt\ (t-t_0)^n
        f(t\mid x_0,v_0)\notag\\ &=n\int_{t_0}^\infty
        dt\ (t-t_0)^{n-1} S(t\mid x_0,v_0,t_0),
 \end{align}
and, in particular, the MFPT $\tau$  reads
 \begin{align}
 	\tau\equiv\langle T \rangle &= \int_{t_0}^\infty dt\ S(t\mid
        x_0,v_0,t_0) \nonumber \\ &= \int_{t_0}^\infty dt
        \int_\mathcal{D} dx dv \ P(x,v,t\mid
        x_0,v_0,t_0).\label{eq:MFPTCL}
 \end{align}
We understand this expression for $\tau$ in the following way. The
averaged first passage time is the sum of all the probabilities to be
in the domain $\mathcal{D}$ at any time $t$ larger than $t_0$. If the
particle is never in $\mathcal{D}$, the integrand vanishes and so does
the MFPT. If, on the other hand, the particle is always in the domain,
the  integral over the probability distribution is normalized to $1$
and the time integral diverges, leading to an infinite MFPT.
  
 Using the adjoint FP equation, it is possible to find an explicit
 solution for the MFPT,
  \begin{eqnarray}
\lefteqn{ 	\mathcal{L}^\dagger_{FP} \tau  = \int_{t_0}^\infty
  \int_\mathcal{D} dx dv\  \mathcal{L}^\dagger_{FP} P(x,v,t \mid
  x_0,v_0,t_0)} \nonumber\\ &&= -\int_\mathcal{D} dx dv P(x,v,t \mid
x_0,v_0,t_0)\Bigr|_{t=t_0}^{\infty} =1,
 \end{eqnarray}
where we assumed that the probability to be in the domain for $t$
going to infinity vanishes, and we used $P(x,v,t_0\mid
x_0,v_0,t_0)=\delta(x-x_0)\delta(v-v_0)$. To find the
mean first passage time, it is sufficient to solve
$\mathcal{L}^\dagger_{FP} \tau=  1$ with the boundary condition
$\tau=0$ on $\partial \mathcal{D}$. Despite the apparent simplicity of
the equation describing the MFPT, the computation turns out to be
rather involved in practice. 

\subsection{Formal equivalence between the MFPT and the
  flux-over-population method}

 A formal relationship between the flux-over-population and the MFPT
 methods has been shown in
 Refs.~\cite{RevModPhys.62.251,PhysRevE.60.R1}. We have learned in the
 previous section that the MFPT $\tau_{\mathcal{D}}(x_0,v_0)$ is
 defined by the equation
 \begin{align}
 	\mathcal{L}^\dagger_{FP}\tau_{\mathcal{D}}(x_0,v_0) & = 1\,,
        & (x_0,v_0)\in \mathcal{D},
 \end{align}
 and the boundary condition $\tau_{\mathcal{D}}(x_0,v_0)=0$ for
 $x_0\in\partial\mathcal{D}$. The Green's function $g(x,v_x\mid
 y,v_y)$ for the FP operator on $\mathcal{D}$ is defined as
 \begin{align}
 	\mathcal{L}_{FP}(x,v_x)g(x,v_x\mid y,v_y) & =
        k\delta(x-y)\delta(v_x-v_y),
\end{align}
for $ (x,v_x)\in \mathcal{D}$,  and
\begin{align}
	g(x,v_x\mid y,v_y) & = 0,&&x\in\partial\mathcal{D}.
 \end{align}
The Green's function might be interpreted as a stationary probability
distribution, since it is a time-independent solution of the FP
equation at every point of the phase space but $(y,v_y)$. This  point
might be seen as an additional point source of strength $k$. Moreover,
the boundary $\mathcal{D}$ acts as a sink. The conservation of
probability implies that the source strength is related to the
probability to be absorbed per unit time, i.e.,
 \begin{align}
 	k & =  \int_{\mathcal{D}} dx
        dv\ \mathcal{L}_{FP}(x,v_x)g(x,v_x\mid y,v_y)   \nonumber
        \\ &=  \int_{\mathcal{\partial D}} dS_i\ J_i(x,v_x\mid y,v_y),
 \end{align}
 where $J_i$ is the probability current density defined from the FP
 equation
 \begin{align}
 	\mathcal{L}_{FP}(x,v_x)g(x,v_x\mid y,v_y) & = \frac{\partial
        }{\partial x} J_x +\frac{\partial }{\partial v_x} J_v.
 \end{align}
 After a multiplication of the Green's function with the MFPT and the
 integration over the domain $\mathcal{D}$, we obtain
 \begin{align}
 	&\int_{\mathcal{D}} dx dv\  \tau_{\mathcal{D}(x,v_x)}
   \mathcal{L}_{FP}(x,v_x)g(x,v_x\mid y,v_y)  \nonumber \\ &  =  k
   \int_{\mathcal{D}} dx dv
   \  \tau_{\mathcal{D}(x,v_x)}\delta(x-y)\delta(v_x-v_y),
 	\end{align}
and
\begin{align}
	& \int_{\mathcal{D}} dx dv\ [
    \mathcal{L}^\dagger_{FP}(x,v_x)\tau_{\mathcal{D}}(x,v_x)]g(x,v_x\mid
  y,v_y)  \nonumber \\ &  =  k  \tau_{\mathcal{D}(y,v_y)}.
 \end{align}
Hence, the MFPT becomes
 \begin{align}
 	 \tau_{\mathcal{D}}(y,v_y) & = \frac{\int_{\mathcal{D}} dx
           dv\ g(x,v_x\mid y,v_y)}{ \int_{\mathcal{\partial D}}
           dS_i\ J_i(x,v_x\mid y,v_y)},
 \end{align}
 which is precisely the inverse of the flux-over-population formula
 for the escape rate~\eqref{eq:fop1dformula},
with a source located at $y$ inside the well. 
 
The escape rate derived with the flux-over-population method is formally equivalent to the inverse of the MFPT. The latter
 provides a simple interpretation of the escape problem. The escape
 time, given by the inverse of the escape rate, is similar to the
 average time needed for a particle to leave a domain. However, the
 MFPT faces some practical difficulties when solving for the rate, in
 particular, beyond the overdamped limit. The flux-over-population
 method is better suited to obtain an analytical solution.


\end{document}